\RequirePackage{snapshot}

\documentclass[11pt]{article}
\usepackage{amsmath}
\usepackage{amssymb}
\usepackage[utf8]{inputenc}
\usepackage{authblk}
\usepackage[square,sort,comma]{natbib}
\usepackage{tikz}
\usepackage{tikz-3dplot}
\usepackage{subcaption}
\usepackage[margin=1in,includefoot]{geometry}
\usepackage[font=small,skip=0pt]{caption}
\usepackage{algorithm}
\usepackage{algpseudocode}
\usepackage{refstyle}
\usepackage{hyperref}
\usepackage{color}
\usepackage[title]{appendix}
\usepackage{etoolbox}
\usepackage{setspace}
\patchcmd{\appendices}{\quad}{: }{}{}

\newcommand{\inputVec}{{\bf P}}

\begin{document}

\title{
Strategic Geosteering Workflow with Uncertainty Quantification and Deep Learning:
\\	A Case Study on the Goliat Field}

\author[1]{Muzammil Hussain Rammay}
\author[2]{Sergey Alyaev}
\author[3]{David Selvåg Larsen}
\author[1]{Reidar Brumer Bratvold}
\author[4]{Craig Saint}

\affil[1]{University of Stavanger, Stavanger, Norway}
\affil[2]{NORCE Norwegian Research Centre, Bergen, Norway}
\affil[3]{Vår Energi, Stavanger, Norway}
\affil[4]{Baker Hughes, Bergen, Norway}

\maketitle

\makeatletter
\def\blfootnote{\gdef\@thefnmark{}\@footnotetext}
\makeatother

\blfootnote{*Corresponding authors: Muzammil Hussain Rammay (muzammil.h.rammay@uis.no) and Sergey Alyaev (saly@norceresearch.no)}
\begin{abstract}

The real-time interpretation of the logging-while-drilling  data allows us to estimate the positions and properties of the geological layers in an anisotropic subsurface environment. 
Robust real-time estimations capturing uncertainty can be very useful for efficient geosteering operations. 
However, the model errors in the prior conceptual geological models and forward simulation of the measurements can be significant factors in the unreliable estimations of the profiles of the geological layers. 
The model errors are specifically pronounced when using a deep-neural-network (DNN) approximation which we use to accelerate and parallelize the simulation of the measurements.
This paper presents a practical workflow consisting of offline and online phases. 
The offline phase includes DNN training and building of an uncertain prior near-well geo-model. 
The online phase uses the flexible iterative ensemble smoother (FlexIES) to perform real-time assimilation of extra-deep electromagnetic data accounting for the model errors in the approximate DNN model.
We demonstrate the proposed workflow on a case study for a historic well in the Goliat Field (Barents Sea).
The median of our probabilistic estimation is on-par with proprietary inversion despite the approximate DNN model and regardless of the number of layers in the chosen prior. 
By estimating the model errors, FlexIES automatically quantifies the uncertainty in the layers' boundaries and resistivities, which is not standard for proprietary inversion.


\end{abstract}

\newcommand{\todo}[1]{{\color{red} TODO: #1}}

\section{Introduction}

Recent advances in logging-while-drilling (LWD) technology allow us to sense the subsurface environment tens of meters away from the well using a suit of extra-deep azimuthal resistivity (EDAR) logs. 
These logs can be inverted in real-time to estimate the geo-layer profile during drilling \citep{simulator2014,seydoux2014full}. However, achieving "strategic" geosteering \citep{arata2016high} requires the integration of the measurements into a subsurface model, which includes the relevant geological uncertainties.

Ensemble-based methods such as the Ensemble Kalman filter and iterative ensemble smoothers present a statistically-consistent Bayesian framework for assimilating data to update an uncertain model \citep{alyaev2019decision}. 
However, their performance relies on thousands of parallel executions of forward models of the obtained measurements \citep{jahani2022ensemble,gji_rammay}. 
The thousands of function evaluations using expensive high-fidelity forward models are computationally expensive in real-time workflows or require special infrastructure \citep{dupuis2015automatic}.   
Recently introduced deep-neural-network (DNN) proxies showed robust approximation for modelling azimuthal EM measurements  \citep{shahriari2020deepforward,kushnir2018real}, including the deepest sensing EDAR measurements \citep{alyaev2021modeling,noordin2022mapping}. 
Most of the computational cost for the DNN can be offloaded to offline training, thus giving superior performance compared to Maxwell's equations solvers during operational use.
real-time statistical data-assimilation workflows with EDAR data required for strategic geosteering \citep{nazainin2021SPWLA,gji_rammay,noordin2022mapping,alyaev2021probabilistic,fossum2022verification}.

Although DNNs bring sub-second model performance, they come with additional unknown model errors.
In realistic scenarios, the negligence of the uncertainties related to the model errors during modelling will result in an unreliable estimation of model parameters and uncertainties \citep{rammay2020flexible,rammay2022calibration,Oliver2018}. Achieving practically usable results for the real-time data assimilation with proxy models requires algorithms that can  automatically account for model errors and detect possible multi-modality.
Recently, flexible iterative ensemble smoother (FlexIES) introduced in \citet{rammay2020flexible} showed good performance in the synthetic tests  for assimilating EDAR data modelled by a DNN where model errors were coming from the inaccuracies in the DNN approximation of the forward model \citep{gji_rammay}.
In a realistic setting, additional model errors will come from inaccuracies in the physical simulation and the mismatch between the chosen geomodel and real, complex geology. 
Moreover, data mismatch can be due to the local minima (local or multiple modes), which might be misinterpreted as data errors by IES-type algorithms \citep{gji_rammay}.

This paper describes a complete FlexIES-based workflow for strategic Bayesian geosteering with steps to account for model errors and multi-modality. It consists of offline and online phases. 
The offline phase takes care of the determination of one or several geological priors consistent with an ensemble method and training a DNN proxy model.
The online phase combines the FlexIES with the DNN model to assimilate the real-time EDAR data.
The workflow is verified on data from a historical operation in the Goliat field and compared to a deterministic inversion delivered by the service company post-job \citep{larsen2015extra}. The objectives of this case study are:

\begin{enumerate}
\item Demonstrate that the DNN from \cite{alyaev2021modeling} can be retrained to handle anisotropic field data and then be used for probabilistic real-time interpretation;
\item Use FlexIES  to estimate probabilistic layered geomodel from field data
automatically accounting for model errors coming from geomodel and the DNN;
\item Compare the FlexIES results with the proprietary deterministic inversion.
\end{enumerate}

The rest of the paper is organised as follows:
Section~(\ref{sec:method}) introduces the workflow and describes the steps in its offline and online phases in more detail. 
The workflow is applied to a reservoir section of a well in the Goliat field using the available historical data, as described in Section~(\ref{sec:results}). The conclusions of this paper are summarised in Section~(\ref{sec:conclusions}).

\section{Workflow}
\label{sec:method}

\begin{figure}[H]
    \centering
    \includegraphics[width=\linewidth]{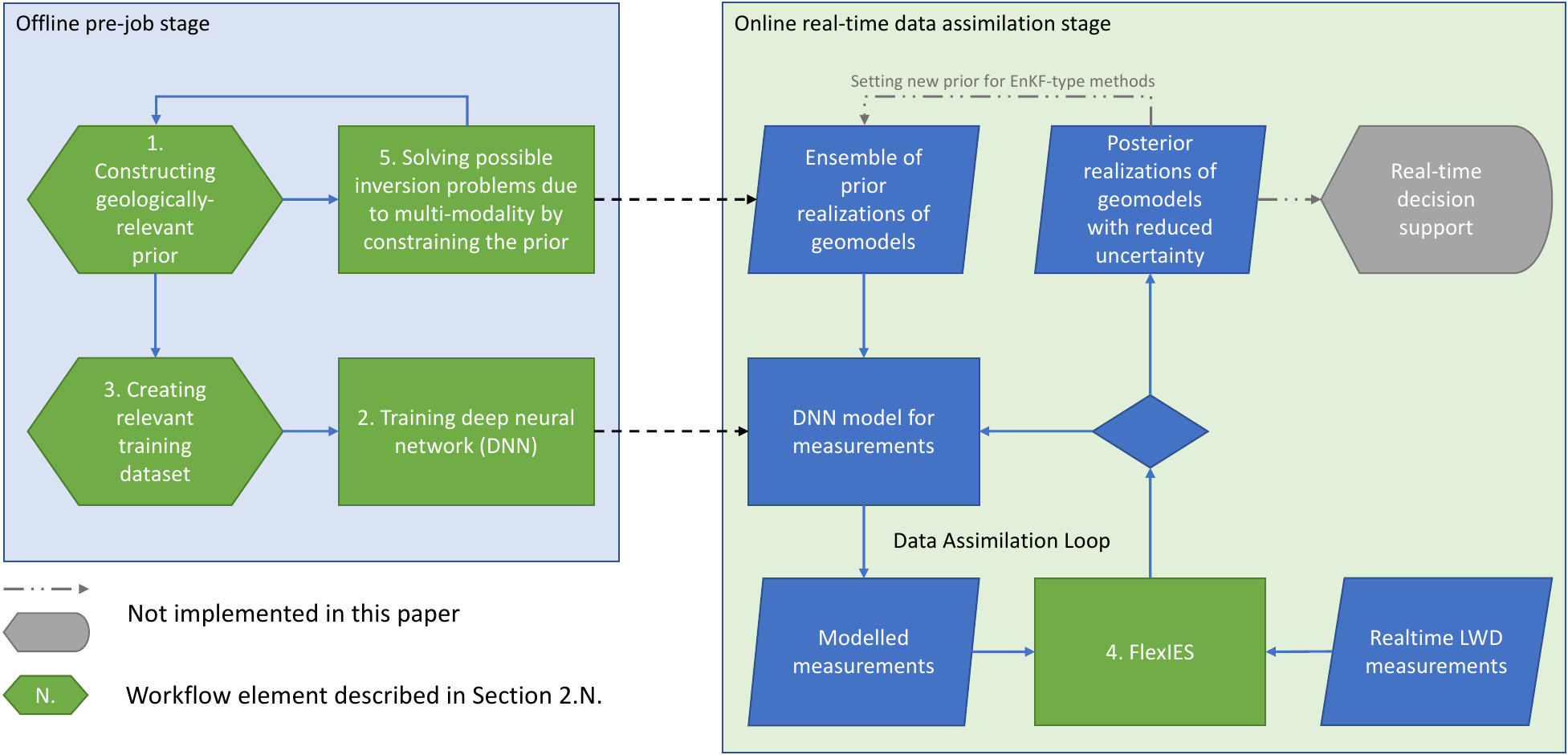}
    \caption{The full workflow for preparation and real-time data assimilation during drilling.}
    \label{fig:fullWorkflow}
\end{figure}

In this paper, we propose an ensemble-based workflow for assimilating borehole electromagnetic measurements to estimate profiles of geo-layers along with associated uncertainties in real-time to support Geosteering operation. 
This workflow consists of offline pre-job stage and an online real-time inversion stage as shown in Figure \ref{fig:fullWorkflow}. 

The offline phase enables the real-time data assimilation.
We start the pre-job phase by constructing a geologically-relevant ensemble of prior geomodel realizations, see Figure \ref{fig:fullWorkflow}.1.
The sampled 1.5D layering configurations from the prior (Figure \ref{fig:fullWorkflow}.3), are used for training a DNN (Figure \ref{fig:fullWorkflow}.2), that can model the suite of EDAR measurements \citep{simulator2014} in milliseconds.
A possibly constrained set of realizations (Figure \ref{fig:fullWorkflow}.5), is used as the prior for the real-time data assimilation loop.

For the online phase we use an Ensemble Kalman Filter (EnKF) type method, namely Flexible Iterative Ensemble Smoother (FlexIES), see Figure \ref{fig:fullWorkflow}.4.
FlexIES compares the synthetic measurements modelled for realizations of the prior ensemble of geomodels with the real-time EDAR measurements in the data assimilation update loop. 
The loop integrates the measurements and reduces the uncertainty in the ensemble yielding the posterior. 
The posterior realizations can be further used for real-time decision support \citep{alyaev2019decision}. 
In case of a filter-type sequential data assimilation the posterior serves as the prior for assimilation of future data \citep{chen2015spe}.

We describe how the workflow is applied to a part of a historical drilling operation, and point out possible modifications needed to apply it for a real operation. 
The rest of the section describe the major building blocks of the workflow with numbering that follows Figure \ref{fig:fullWorkflow}.

\subsection{Constructing geologically-relevant prior}
\label{sec:prior}
The first step of the workflow is related to the prior description of the geo-model. 
This can be done by utilizing the prior knowledge or experience of the geologists or the interpretation of the data set from the offset wells. 
In the considered historical case the geology can be represented by a layer-cake model with continuous layers \cite{larsen2015extra}.
Thus, we describe 
the geo-model by a number of layers, the 2D profile of each layer, and the anisotropic layer's resistivities. 
For example, three layers geo-model and four layers geo-model are shown in Figure \ref{syn_truth}. 

The prior uncertainties of a geo-model are described by prior realizations of the geo-layer profiles (estimated as boundary positions or layers thicknesses) and layers' resistivities. 
We model the prior realizations of the thickness profiles of the geo-layers as multivariate Gaussian with mean ($\mathbf{\mu} = [35, 30, 20]$) and an exponential covariance function:
\begin{equation}
\mathbf{c} = \sigma^2 \exp{(-3\frac{\mathbf{h}}{l})}
\end{equation}
where $\mathbf{h}$ is the lag distance and $\sigma^2$, $l$ are the variance and  correlation length respectively (which are 33.78 and 20 in this work). 
Thus our prior geo-model realizations do not have any trends on layer dips as shown with the green lines in Figure \ref{est_1b}. The thickness profiles of the geo-layers are converted to the boundary positions of the layers by adding the fixed top position with the thickness of the geo-layers.  
Furthermore, the prior realizations of the logarithm of the layer resistivities are independently sampled from the normal distributions for the sand and the shale layers. The means of log-resistivities of the shale and sand layers are $0.5$ and $2.5$ and standard deviations are $0.25$ and $0.5$ respectively.

\subsection{Forward DNN of EDAR measurements: description and training}

The physical forward model for EDAR measurements can be represented by Maxwells' equations \cite{luo2015ensemble}. 
The measurements come from the values of components of the magnetic field induced by the tool's transmitters evaluated at the tool's receivers. 
In practice, EDAR log values include several frequencies and are corrected for environmental factors and compensated using extra receivers \cite{alyaev2021modeling}.
Formally, the EDAR forward model can be written as a vector function of the subsurface properties:

\begin{equation}
		\mathbf{EM} = F\mathbf{(S,t)},
		\label{eq:physics}
\end{equation}
where $F$ is the function which maps subsurface properties $\mathbf{S}$ (e.g., layers positions/boundaries, resistivities) to electromagnetic $\mathbf{EM}$ signals for a given well trajectory $\mathbf{t}$.

Any forward model is to some extent low-fidelity approximation of the reality.
In this work, we use a DNN approximation of a vendor-provided EDAR model \cite{simulator2014}, and try to account for its higher inaccuracy using FlexIES \cite{gji_rammay}. The approximation can be formally written as:
\begin{equation}
		\mathbf{\widetilde{EM}} = \mathbf{F}(\mathbf{S,t}) = DNN(\inputVec),
\label{dnn_eq1}		
\end{equation}
where $DNN$ is the DNN approximation of the EDAR EM signals, and $\inputVec$ is the input vector representing the local subsurface configuration and relative well trajectory.
In this work we use the structure of the DNN described in \cite{alyaev2021modeling}.
The inputs and the outputs of the DNN are the same as in the vendor EDAR software. 
The structure of the DNN is summarized in the following subsections together with the training procedure.

\subsubsection{DNN inputs}
\label{sec:DNNinputs}
To simplify solution of Maxwells' equations, vendor models often assume layer-cake geology sampled near the tool location. 
We encode seven subsurface layers around the current tool position. This setup, three layers above and three layers below the logging position layer, is close to the practical tool's best-case look around capability.
Thus, for each logging point we get the following inputs:
\begin{itemize}
    \item six boundary positions between layers, encoded in meters relative to the tool position (the top and the bottom layers are considered semi-infinite);
    \item seven pairs of anisotropic resistivities for each of the layers: parallel and perpendicular to the layer;
    \item two angles determining the geometry of the well relative to the layers: one recorded at the logging point, and one 20 m ahead of the logging position.
\end{itemize}
In total, this sums up to 22 input variables that we denote as vector $\inputVec \in \mathbb{R}^{22}$.

\subsubsection{DNN outputs / EDAR logs}
\label{sec:logTypess}

\begin{table}
    \centering
\includegraphics[width=\columnwidth]{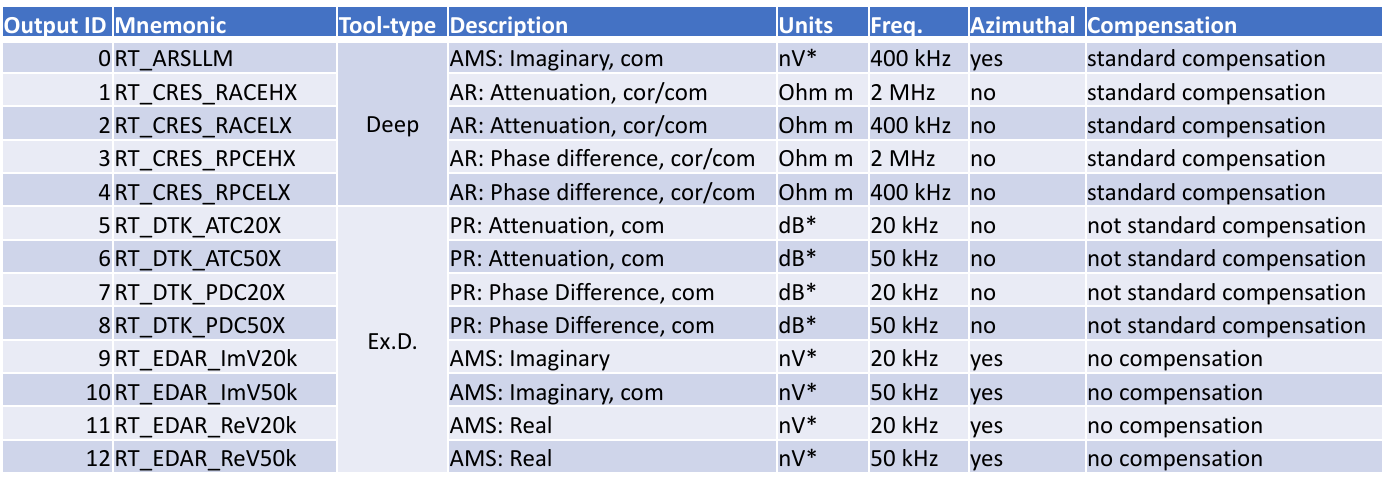}
    \includegraphics[width=0.7\columnwidth]{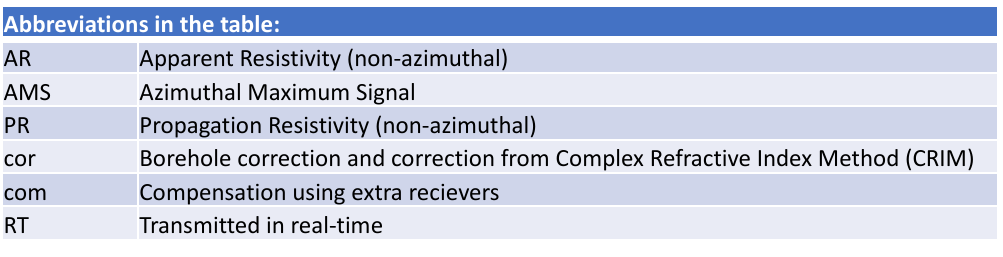}
    \caption{Output IDs, mnemonics, and descriptions of the log signals used in the paper. 
}
    \label{tab:mnemonics}
\end{table}

The DNN approximates the full EDAR suite of 22 logs used during a real-time operation:  four shallow apparent resistivities 
and nine pairs of deep directional measurements.
We adopt the 2D earth model assumption (i.e. assume there is no azimuth, sideways, angle), which allows to replace the directional angle-value pairs by a single signed number.
This gives a total of 13 signed output values, which we denote by $\mathbf{O} \in \mathbb{R}^{13}$. 
Table~\ref{tab:mnemonics} summarises the mnemonics of the signals and corresponding DNN output IDs used in the numerical results. 
For detailed description of the logs, we refer the reader to \cite{alyaev2021modeling}.
In the numerical experiments the signal magnitudes are normalized by fitting the scikit-learn's MinMax scaler \citep{scikit-learn} with range $0.5..1.5$ for the prior realizations of the outputs.

\newcommand{\scalerIn}{\Psi_{in}}
\newcommand{\scalerOut}{\Psi_{out}}

\subsubsection{DNN architecture and training}
\begin{figure}
    \centering
    \includegraphics[width=0.7\columnwidth]{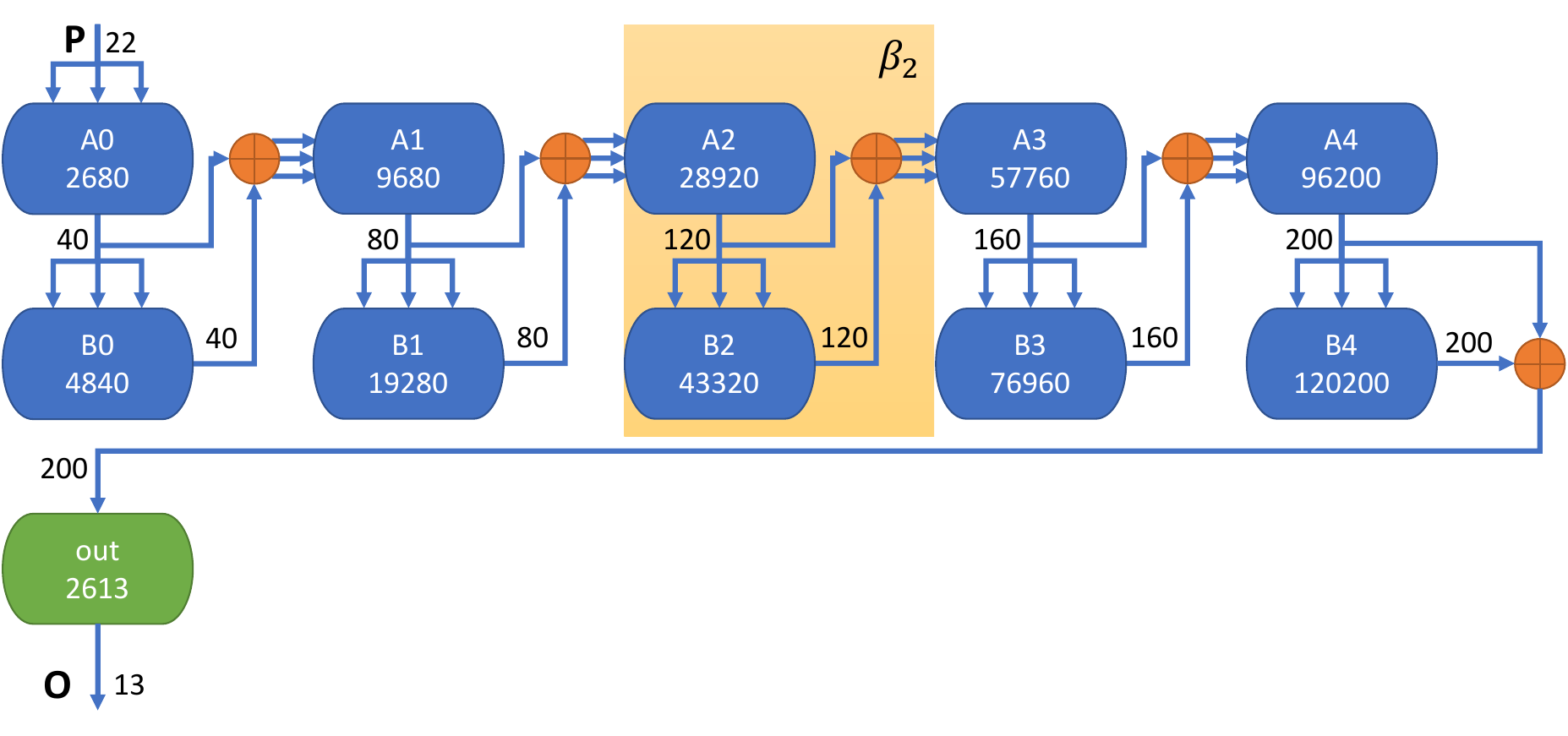}
    \caption{The schematics of the DNN architecture. 
    The DNN is composed of five consecutive fully-connected layer pairs ($Ai$, $Bi$), each having ReLU activation. 
    The outputs of the layer pairs are summed together, forming structure similar to residual blocks, denoted $\beta_i$. 
    The pairs are followed by the linear output layer.
    The numbers in each box are the sizes trainable parameters, and the numbers on the arrows are the sizes of inputs / outputs of layers.
}
    \label{fig:architecture}
\end{figure}

We adopt the DNN architecture from \cite{alyaev2021modeling}, which consists of five consecutive convolutional residual blocks followed by a linear output layer, see Figure \ref{fig:architecture}. 
Each convolutional layer is degenerate, that is the input is repeated three times and then convolved with a convolutional filter of size three, translating into a redundant fully-connected structure, see Figure \ref{fig:architecture}.
The $DNN$ function can be written as
\begin{equation}
        DNN(\inputVec) = \scalerOut^{-1} 
        \circ \textrm{out}_{w_{5}}  \circ
        \beta^{4}_{w_{4}} \circ 
        \beta^{3}_{w_{3}} \circ \cdots \circ \beta^{0}_{w_{0}} \circ
        \scalerIn \inputVec,
\end{equation}
where $\circ$ is the function composition; the $\scalerIn$ and the $\scalerOut$ are the scikit-learn's MinMax scalers \citep{scikit-learn} fitted to the training data; and $\beta^{i}$ and '$\textrm{out}$' are neural network structure blocks with trainable-parameter weights $w_{0}$.
The block architecture and the numbers of trainable parameters are illustrated in Figure \ref{fig:architecture}. 
Given a relatively large model size of 462,453 trainable parameters, it is expected to generalize to different geological settings using unmodified training procedure \citep{alyaev2021modeling}.

A synthetic dataset is split 80 / 10 / 10 into training validation and testing data respectively. 
During training we minimize the Mean Absolute Error loss using the default settings of the Adam optimizer \citep{kingma2014adam} implemented in TensorFlow \citep{tensorflow2015whitepaper} with the batch size of 512.
The validation data is used for the Tensorflow's early stopping.
The testing data are only used to assess under- or over-fitting after the training is completed, see Appendix \ref{sec:trainingResults}. 
The architecture and the training setup described above are exactly the same as in \cite{alyaev2021modeling}. In this paper, however, we train the DNN on a larger and more complex data set described in Section \ref{sec:dataset} to verify the robustness of the DNN setup and to make the DNN model applicable to field cases.

\subsection{Creating relevant training dataset}
\label{sec:dataset}

An accurate model of the measurements is needed to provide the best interpretation results from data assimilation. 
Even though the data-driven models inherently contain model errors, we want to make sure that they resemble the physics of the measurements in the best possible way.
In our workflow we use a vendor-provided forward model which incorporates a number of corrections, see Table \ref{tab:mnemonics}, to produce a large synthetic dataset offline.
Due to common license limitations on simulation software a graphical user interface automation script AutoIt \cite{autoit} was used to create the samples, making this offline stage relatively expensive in terms of computational time.

\subsubsection{Conversion of geo-model to the DNN input}
Companies put effort into creating likely subsurface scenarios that may be observed during drilling. 
These scenarios are the basis for generating the prior as we describe in Section \ref{sec:prior}. 
They can be sampled with different relative well angles and positions to emulate likely configurations of the number of layers, thickness, and resistivity ranges. 

We implement the following procedure to convert the near well geological realizations to the DNN inputs as shown in section \ref{sec:DNNinputs}. The realization of a geological model is divided into seven layers using the current position of the logging tool $t_p$, and boundary positions of the geological layers $l_p$. We compute the layer number of the tool location in the geological model by comparing the current position of the well with the upper and lower boundary positions of all geological layers. After that, we assign DNN inputs by computing six boundary positions using $b_p = l_p-t_p$ and taking fourteen resistivities (7 pairs of anisotropic resistivities) above and below the tool position as shown in Figure \ref{fig:layered}. There is a possibility that in a given geo-model there are less than seven layers or the tool is in the first or last layer. In that situation, we subtract and add 10 to the existing upper and lower boundary positions respectively in order to obtain six boundary positions. Furthermore, if there are less than seven layers in the geo-model, the missing pairs of anisotropic resistivities are taken from the existing pairs of anisotropic resistivities of the first and last layers.

\begin{figure}
\begin{center}
a.
\includegraphics[width=0.45\textwidth]{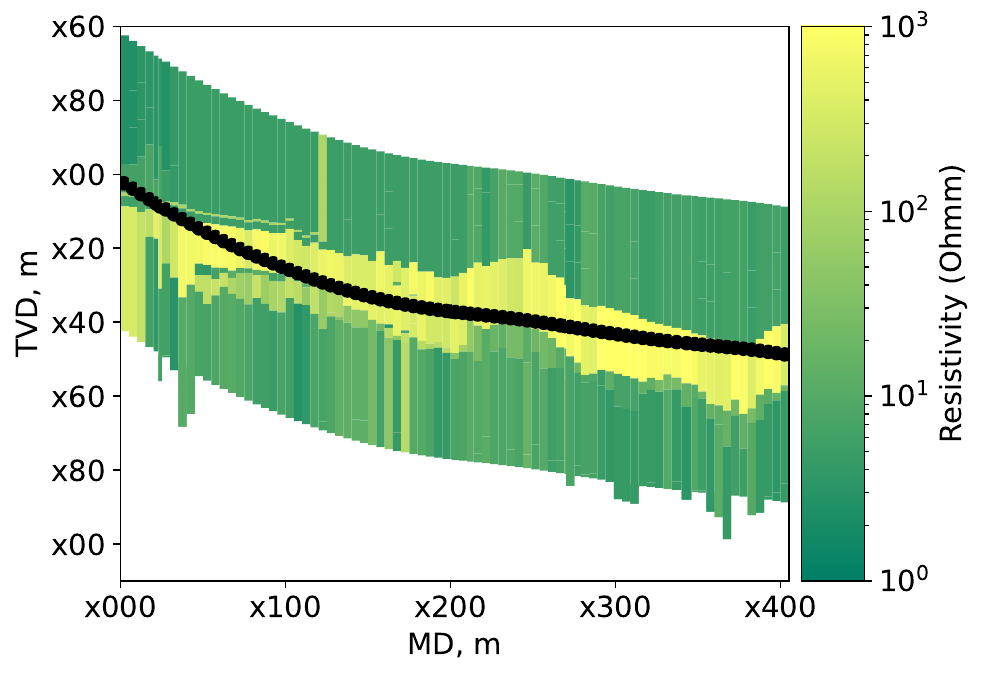}
b.
\includegraphics[width=0.45\textwidth]{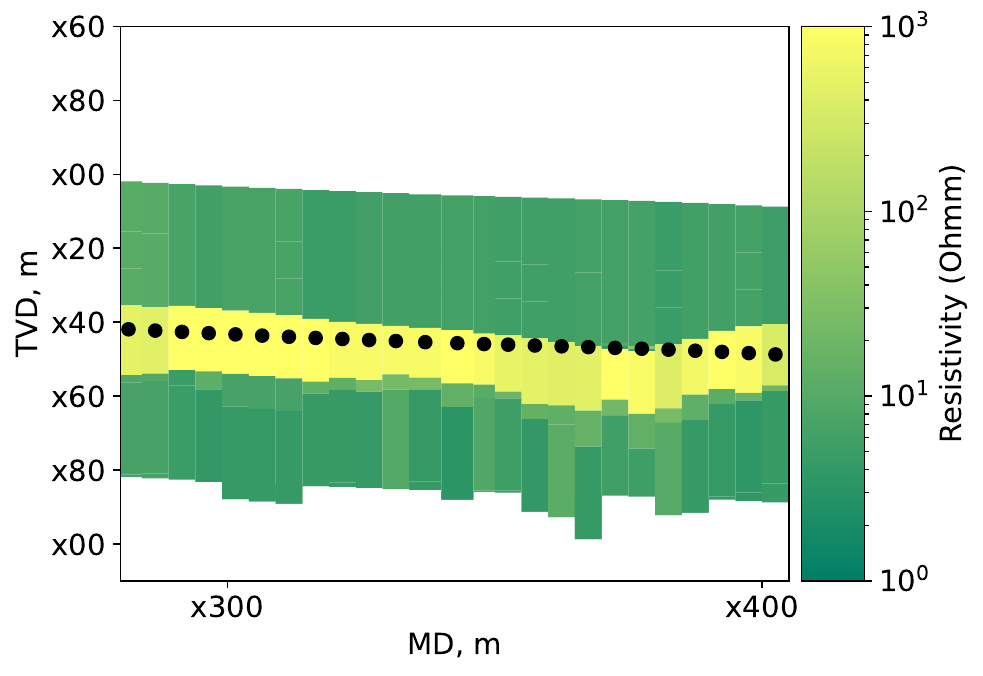}
\end{center}
\caption{a. The reference vendor-provided inversion described in \cite{larsen2015extra}. 
b. The case study section of the well for this paper. Dots show the well position in the middle of the inversion intervals along the well trajectory.
}
\label{prop_inversion}
\label{fig:proprietary}
\end{figure}

\subsubsection{Generation of the dataset for a historical case.}
In the current paper, we use a simplified setup, taking the advantage of the available proprietary inversion.
The subsurface seven-layer configurations were generated from a random distribution based on the available proprietary inversion, see Figure \ref{fig:proprietary}. 
The inversion consists of a number of 1D inversion slices, each having a different number of layers, with different anisotropic resistivities and inclination with respect to the well.

\begin{figure}
    \centering
    \includegraphics[height=0.33\columnwidth]{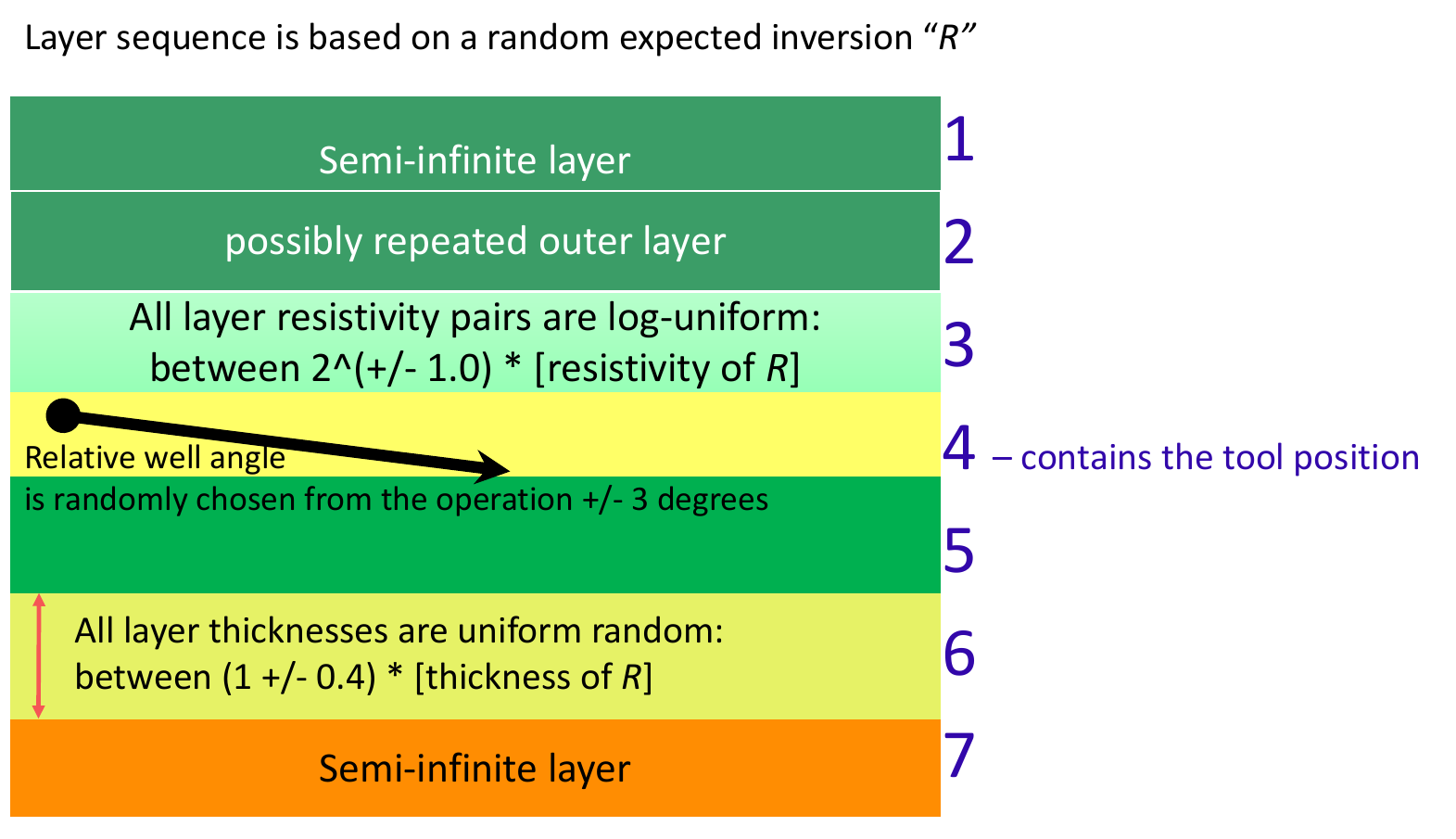}
    \caption{
    Schematics of 1D input from a training dataset for the DNN forward model.}
    \label{fig:layered}
\end{figure}

The 1D data samples were generated as follows, see Figure \ref{fig:layered}:
\begin{itemize}
    \item A random configuration from the 1D inversion slices based on the operation was randomly selected. The number of layers was used from the selected configuration.
    \item Each layer was assigned an anisotropic resistivity, with each component in log-uniform range from $\rho^*/2$ to $\rho^*2$, where  $\rho^*$ is the value of resistivity component for this layer in the selected configuration.
    \item The boundary positions were perturbed randomly, ensuring that the boundary sequence is kept consistent. The thicknesses were distributed uniformly in range from $0.6 * h^*$ to $1.4 * h^*$, where $h^*$ is the thickness of a given layer.
    \item If the configuration had less than seven layers, the outer layers on the top and the bottom were repeated every 10 m to yield seven-layer model.
    \item The relative well angle was selected on-random separately from the historical inversion and perturbed by +/- 3 degrees.
\end{itemize}
Using the described procedure, we generated a dataset consisting of around 1.1 million inputs.
For each of the samples, we run the forward model for all 22 logs without adding noise.
The resulting dataset is subdivided into 80/10/10 for training  validation and testing as described above. 
The training for the presented case converged after 338 epochs, yielding a coefficient of determination of 0.97 and above for the synthetic test data.
The summary of the training can be found in Appendix \ref{sec:trainingResults}.
In the numerical results, we will use the trained network without further adjustment.

\subsection{Flexible iterative ensemble smoother (FlexIES)}\label{subsec:FES}

Extensive literature reviews about the algorithms to account for model-error can be found in the following papers; 
1) utilization of known prior model-error statistics computed from pairs of high-fidelity and low-fidelity models \citep{rammay2019quantification,rammay2020robust}, 
2) addressing complex error statistics using orthogonal basis generated from pairs of high-fidelity and low-fidelity models \citep{Corinna2019,KOPKE2017}, 
3) estimation of unknown model-error from the residual (data mismatch) during the inversion \citep{Oliver2018,rammay2020flexible}.  

In this workflow, we use FlexIES \citep{rammay2020flexible} as a Bayesian inversion algorithm to invert the EDAR logs in order to obtain real-time estimations of the geo-layers profile along with layers' resistivities. In real-time inversion, ignoring model errors could provide unreliable estimate of the geo-layers profiles. In that situations, FlexIES can be useful to provide more reliable estimates of the geo-layers profile and prohibit to converge to the wrong solution. FlexIES estimates the ensemble of the mode errors during the inversion using the ensemble of data-mismatch/residual and split parameter $s_p$. In the algorithm, the split parameter is computed from the ratio of norm of mean residual in the current and previous iterations as shown in Appendix \ref{sec:AppendixA}. FlexIES can provide exact data-match if there is no model-errors in the prior geo-models or measurement errors in the logging instruments. In order to compare the results with the inversion while neglecting model-errors, we use ensemble smoother with multiple data assimilation (ESMDA) algorithm \citep{emerick2013}.

\label{sec:formulation}

The DNN and FlexIES are used in a combined framework in order to perform the inversion in real-time. Utilization of the DNN model allows us to perform Bayesian inversion of the EDAR logs in real time, which is useful to obtain real-time estimations of the geo-layers profile along with layers' resistivities. Moreover, we can also quantify the non-uniqueness and uncertainties related to the prior geo-models in real-time. For this purpose, we perform the real-time inversion using iterative ensemble smoothers due to their computational efficiency, flexibility for highly non-linear models and parallel nature of the algorithms.

The framework of the FlexIES for real-time inversion of EM measurements while utilizing DNN as a forward model are shown in Figure~(\ref{esmda_diag}) and detailed algorithm in Appendix \ref{sec:AppendixA}. The first step is related to the computation of the $N_e$ number of the prior realizations (ensemble members) of $N_m$ number of the geo-layers profiles (estimated as layers thicknesses or boundary positions) and layers resistivities $\mathbf{S}$ from a known or assumed statistical distribution. In this work, the prior realizations $\mathbf{M}_{prior} = [\mathbf{S}_{1} \; \mathbf{S}_{2} \; \mathbf{S}_{3} \; ...... \; \mathbf{S}_{N_e}]$  of the geo-layers profiles and anisotropic resistivities are taken from the selected prior geo-model in the offline stage.

In the next step, the realizations of the outputs $\mathbf{D}$ {(\color{black} ensemble of the Electromagnetic signals)} are computed by passing the prior realizations of the profiles of geo-layers and resistivities to the DNN for a given well trajectory $\mathbf{t}$ i.e.~$\mathbf{D} = DNN(\mathbf{M, t}) \in [\mathbf{\widetilde{EM}}_{1} \; \mathbf{\widetilde{EM}}_{2} \; \mathbf{\widetilde{EM}}_{3} \; ...... \; \mathbf{\widetilde{EM}}_{N_e}]$. {\color{black} The prior ensemble of the geo-layers profile and electromagnetic outputs are used to estimate the posterior ensemble of the geo-layers profile and resistivities using the FlexIES algorithm} for the observed electromagnetic measurements $\mathbf{EM}_{obs}$ of size $N_d$:
\begin{equation}
    \mathbf{M}_{post}, \mathbf{D}_{post} \gets {FlexIES}(\mathbf{M}_{prior}, \mathbf{EM}_{obs}, \mathbf{C}_D, DNN, \mathbf{t}),
    \label{alg:A1}
\end{equation}
where $\mathbf{C}_D$ is the covariance of the measurement error.

\begin{figure}
\begin{center}
\includegraphics[width=1\linewidth]{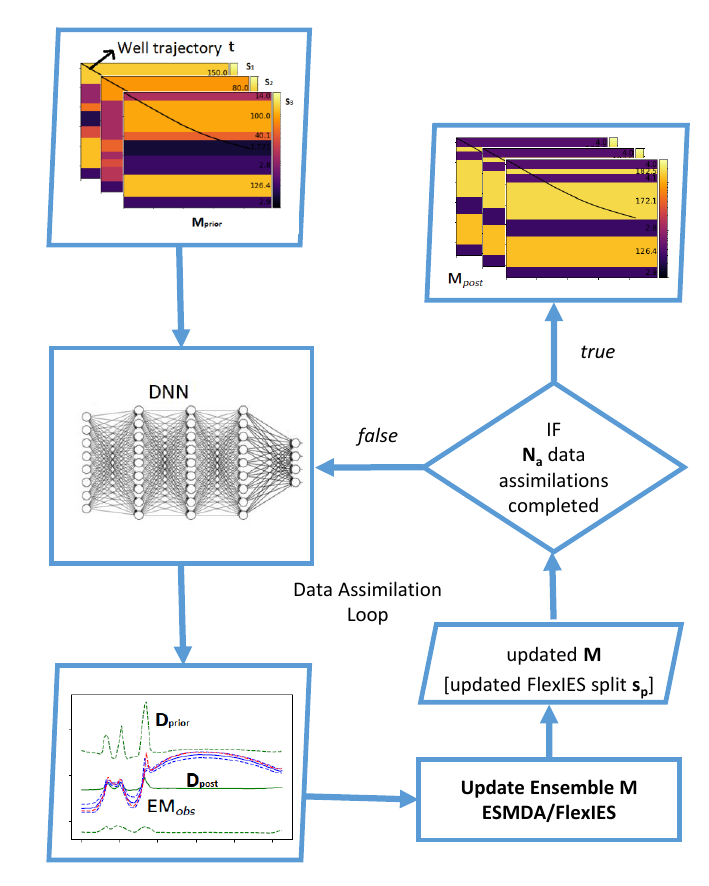}
\end{center}
\caption{Schematic diagram of FlexIES algorithm for real-time inversion of EM measurements by utilizing DNN as a forward model.}
\label{esmda_diag}
\end{figure}

\subsection{Identification of multi-modality and non-uniqueness in prior geo-models} \label{sec:identification}

Practically we are uncertain about the extent of the multi-modality, non-uniqueness and models imperfection. In these scenarios, the identification of the multi-modality and non-uniqueness are required in the selected prior geo-models before the realistic inversion. This step should be done in the absence of the model-error and the measurement errors, otherwise, it would be difficult to distinguish the reason of the data-mismatch. For this purpose, we generate the observations from the same DNN model by considering a realization of the prior geo-model as a synthetic truth. If there is no multi-modality or non-uniqueness problem, we should get an exact data-match and exact estimations of the synthetic truth after the inversion.

\begin{figure}
\begin{center}

 \hspace{-0.7in}
    \begin{subfigure}[normal]{0.4\textwidth}
	\includegraphics[scale=0.5]{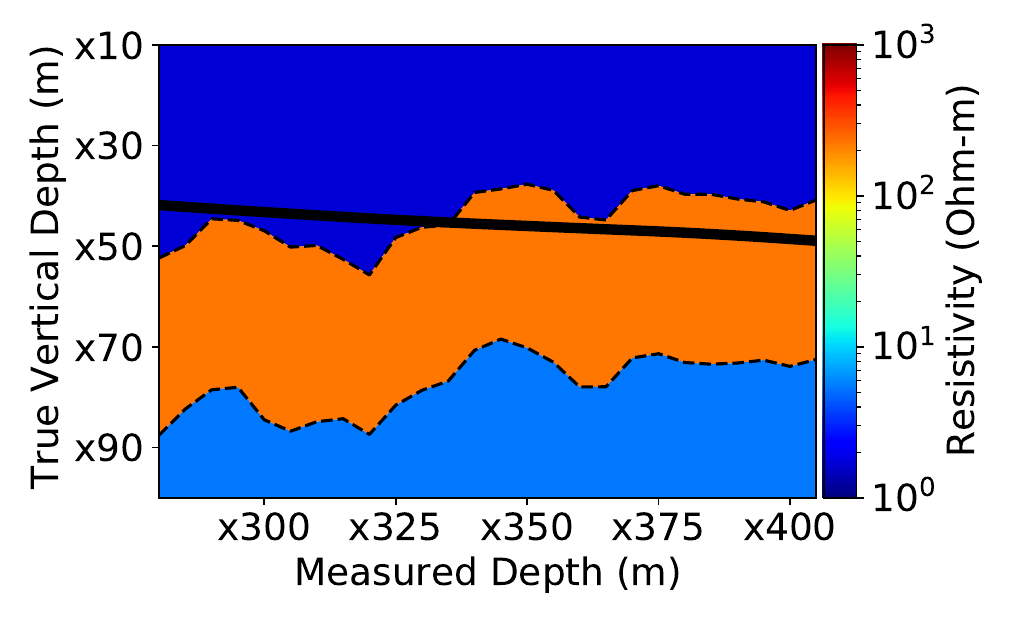}	
	\end{subfigure}
 \hspace{0.45in}
	\begin{subfigure}[normal]{0.4\textwidth}
	\includegraphics[scale=0.5]{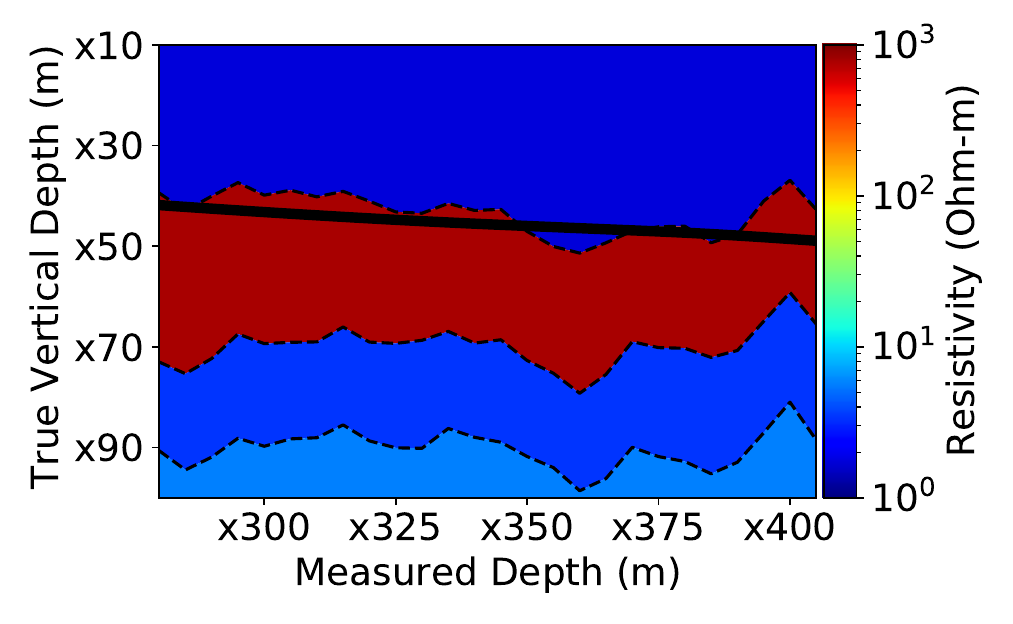}	
	\end{subfigure}

 \hspace{0.1in}
		 Three layers geo-model
 \hspace{1.5in}
 		 Four layers geo-model 		 
 		 
\end{center} 
 
\caption{Synthetic reference models for multi-modality identification: three and four layer prior geo-models.}
\label{syn_truth}
\end{figure}

\begin{figure}
\begin{center}

 \hspace{-0.5in}
    \begin{subfigure}[normal]{0.4\textwidth}
	\includegraphics[scale=0.5]{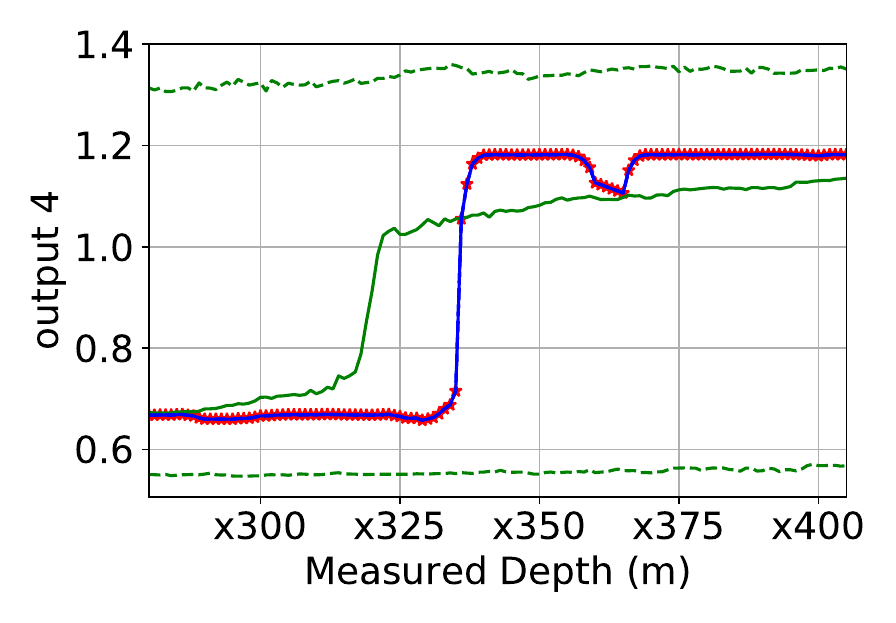}	
	\end{subfigure}
 \hspace{0.5in}
	\begin{subfigure}[normal]{0.4\textwidth}
	\includegraphics[scale=0.5]{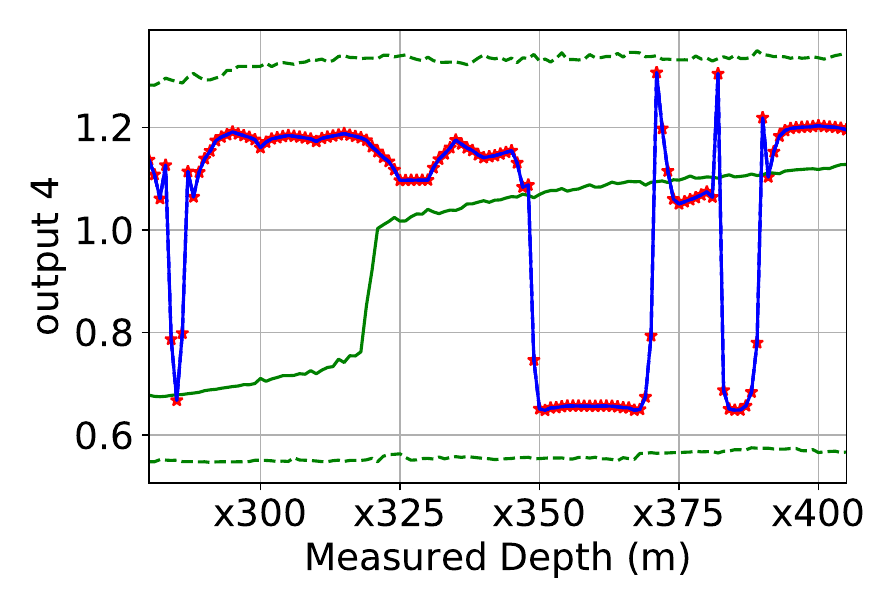}	
	\end{subfigure}

 \hspace{-0.5in}
    \begin{subfigure}[normal]{0.4\textwidth}
	\includegraphics[scale=0.5]{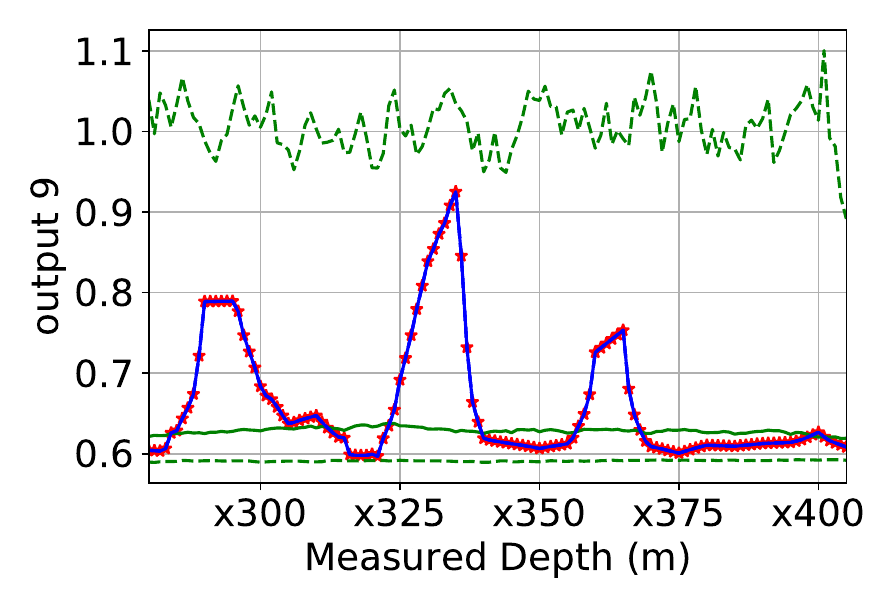}	
	\end{subfigure}
 \hspace{0.5in}
	\begin{subfigure}[normal]{0.4\textwidth}
	\includegraphics[scale=0.5]{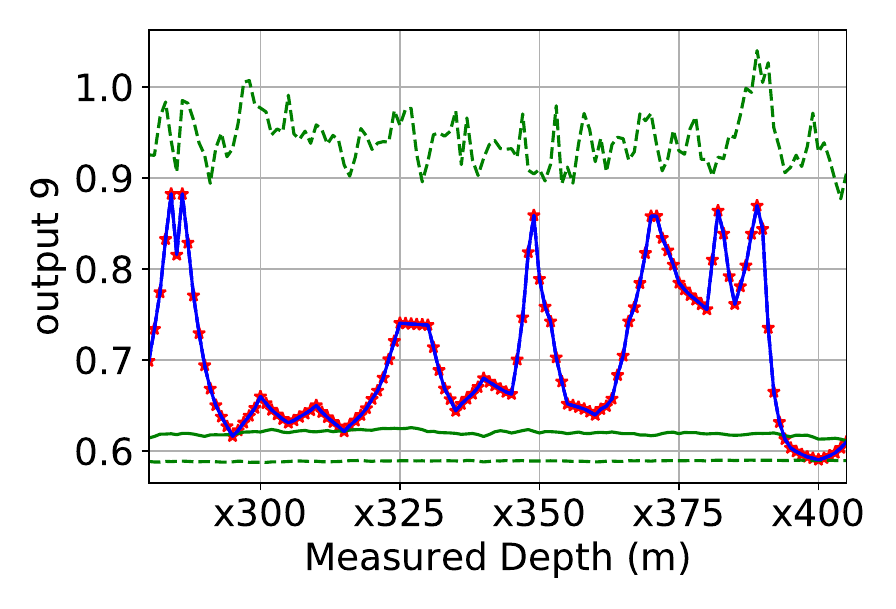}	
	\end{subfigure}

 \hspace{-0.5in}
    \begin{subfigure}[normal]{0.4\textwidth}
	\includegraphics[scale=0.5]{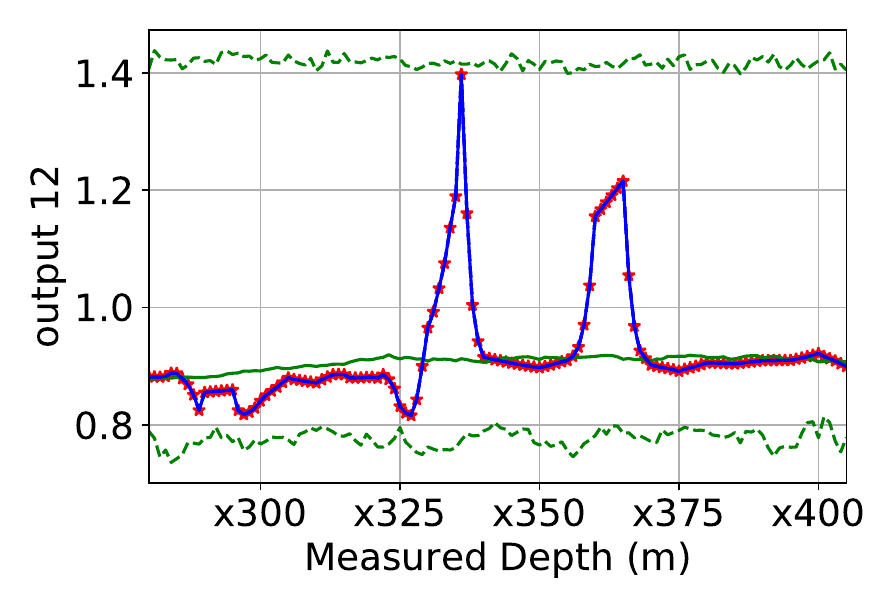}	
	\end{subfigure}
 \hspace{0.5in}
	\begin{subfigure}[normal]{0.4\textwidth}
	\includegraphics[scale=0.5]{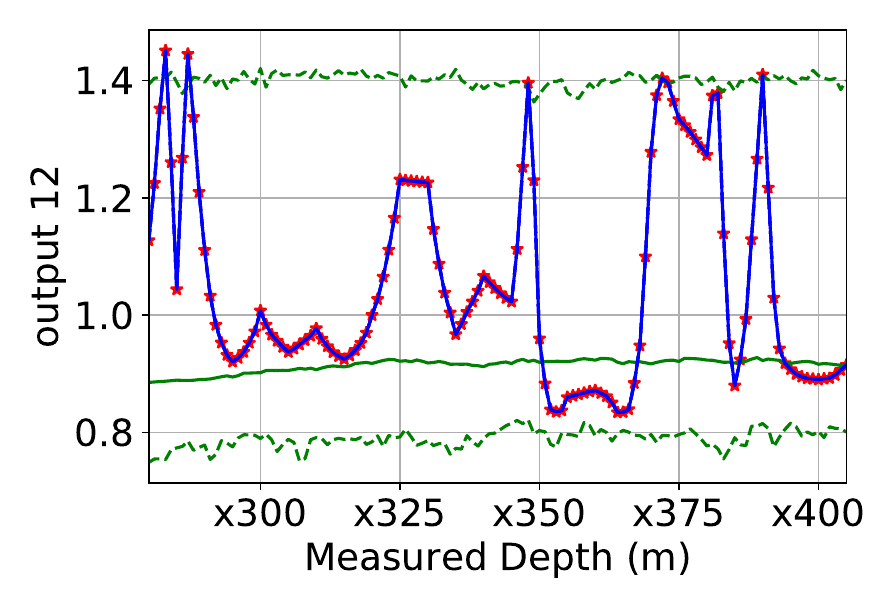}	
	\end{subfigure}

 \hspace{0.5in}
		 Three layers geo-model
 \hspace{1.5in}
 		 Four layers geo-models
 \end{center}

\caption{Prior and posterior distribution of EM outputs considering DNN as a perfect forward model. Green and blue lines show ensemble approximation of the prior and posterior distribution respectively and red stars show observed EM measurements. Solid green and solid blue lines show 50th percentile ($p50$) and dashed green and dashed blue lines show $98\%$ confidence interval respectively of the prior and posterior distribution respectively. The posterior distribution appears as the point estimate therefore solid blue lines overlaps dashed blue lines. First column of the sub-figures show results obtained from the three layers geo-model and second column of the sub-figures show results from the four layers geo-model.}
\label{est_outputs_1}
\end{figure}

There is a possibility that our prior geo-models are not realistic in terms of the description of the number of geo-layers. Therefore we generate two synthetic truths from realizations of two prior geo-models of three and four layers as shown in Figure \ref{syn_truth}. Figure \ref{est_outputs_1} shows the exact data match of the EM measurements thus shows no multi-modality or non-uniqueness issue in the physical simulation of the selected prior geo-models of the three and four layers.  

\begin{figure}
\begin{center}

 \hspace{-0.7in}
    \begin{subfigure}[normal]{0.4\textwidth}
	\includegraphics[scale=0.5]{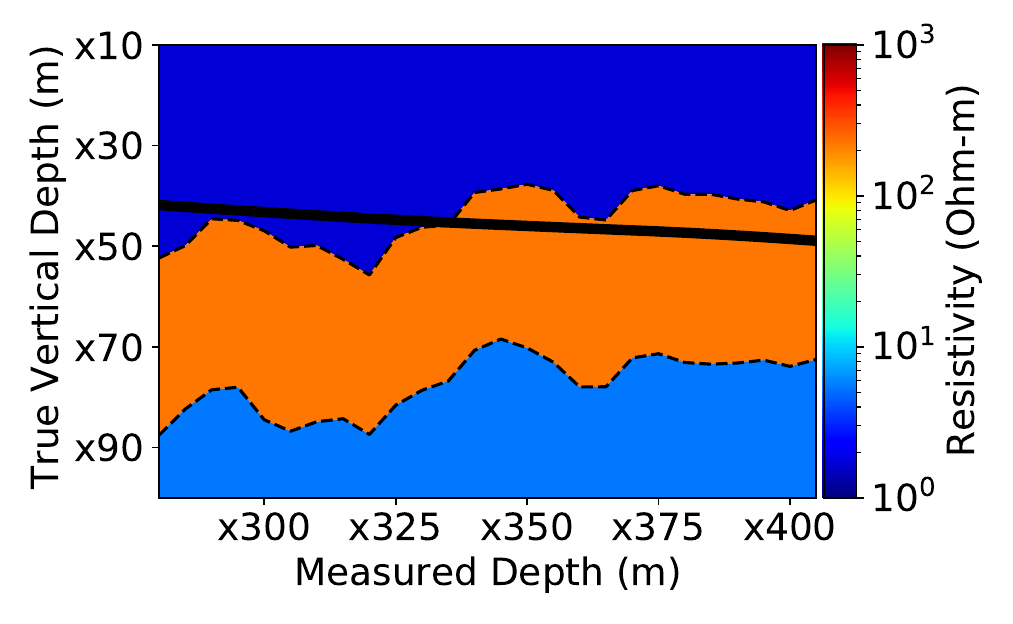}	
	\end{subfigure}
 \hspace{0.45in}
	\begin{subfigure}[normal]{0.4\textwidth}
	\includegraphics[scale=0.5]{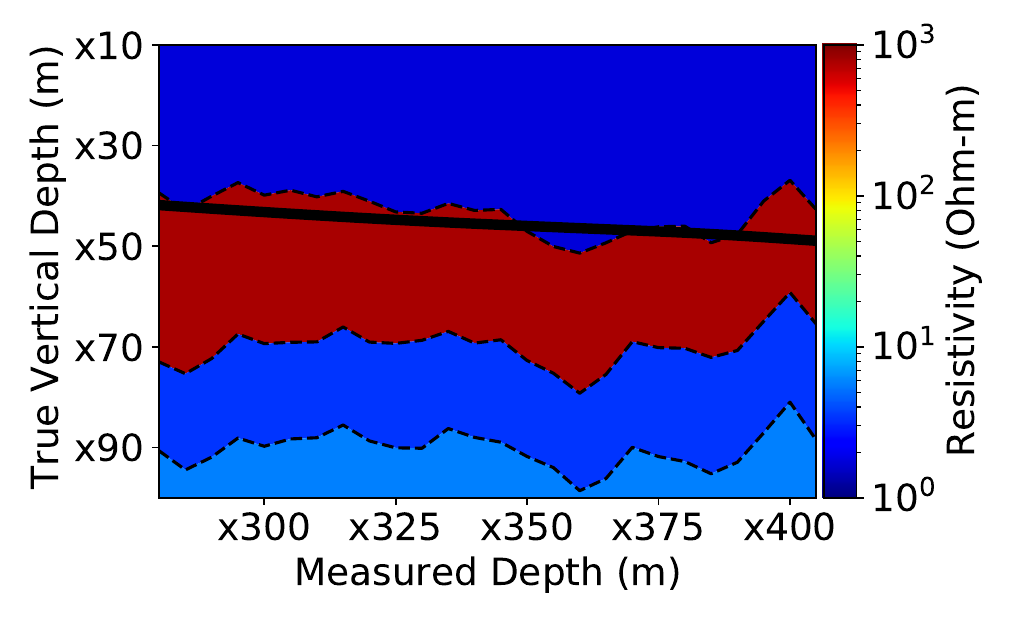}	
	\end{subfigure}

 \hspace{0.1in}
		 Three layers geo-model
 \hspace{1.5in}
 		 Four layers geo-models
 \end{center}

\caption{Estimation results of the geo-layers profile for identification of multi-modality and non-uniquesness in prior geo-models.}
\label{est_1a}
\end{figure}

\begin{figure}
\begin{center}

 \hspace{-0.5in}
    \begin{subfigure}[normal]{0.4\textwidth}
	\includegraphics[scale=0.5]{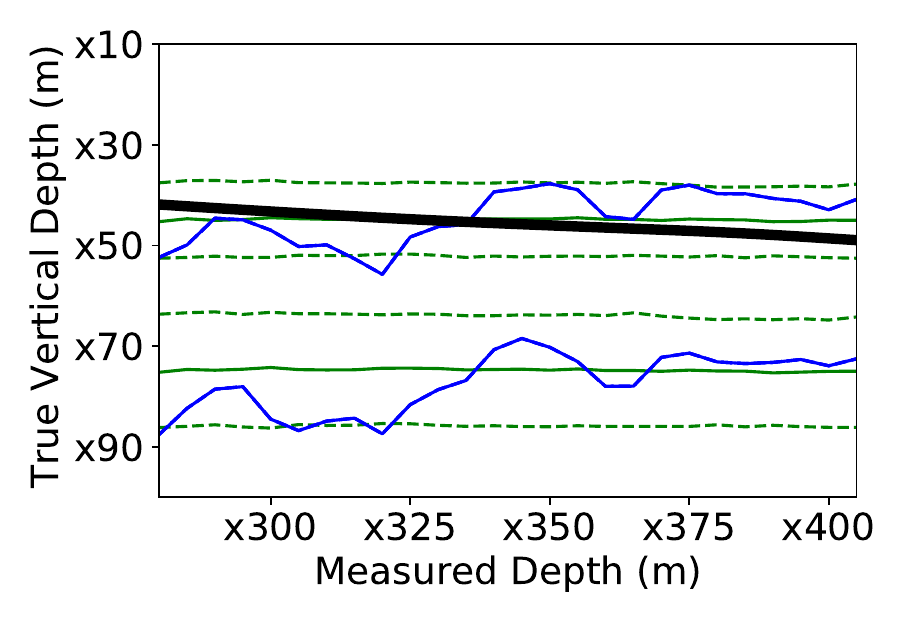}	
	\end{subfigure}
 \hspace{0.5in}
	\begin{subfigure}[normal]{0.4\textwidth}
	\includegraphics[scale=0.5]{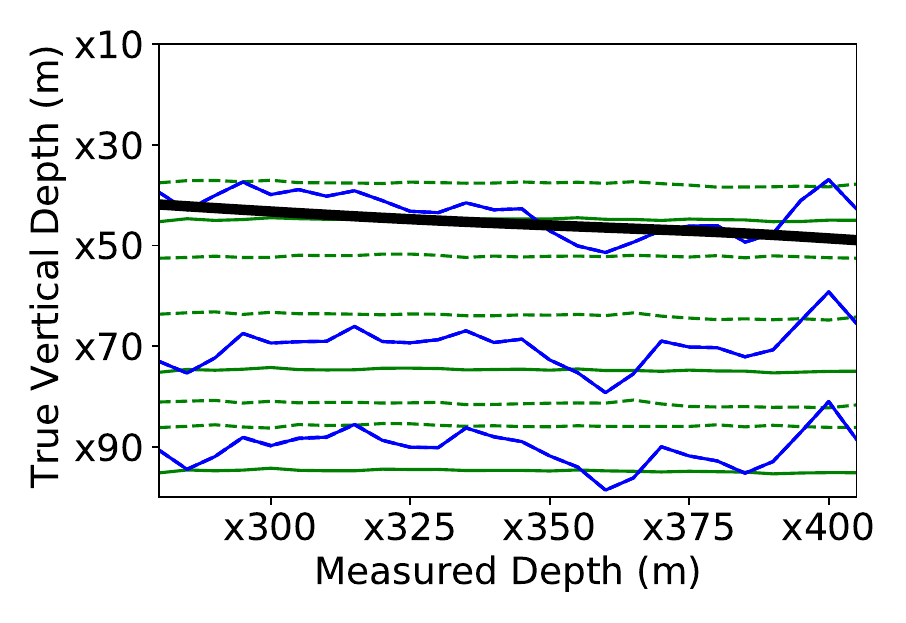}	
	\end{subfigure}

 \hspace{0.5in}
		 Three layers geo-model
 \hspace{1.5in}
 		 Four layers geo-models
 \end{center}

\caption{Prior and posterior distribution of the geo-layers profiles. Green and blue lines show ensemble approximation of the prior and posterior distribution respectively. Solid green and solid blue lines show 50th percentile ($p50$) and dashed green and dashed blue lines show $80\%$ confidence interval respectively of the prior and posterior distribution respectively. The posterior distribution appears as the point estimate therefore solid blue lines overlaps dashed blue lines.}
\label{est_1b}
\end{figure}

Figure \ref{est_1a} shows the estimation results of the geo-layers profiles and resistivities for both three and four layers models. These estimation results  are exactly same as the synthetic truths as shown in Figure \ref{syn_truth}, thus showing no uncertainties related to the non-uniqueness and multi-modality. Furthermore, the posterior distributions appear as a point estimate as shown in Figure \ref{est_1b}. These results confirm the absence of multi-modality and non-uniqueness in the selected prior geo-models of three and four layers.

\section{Realistic inversion and estimation of the geo-layers profiles of the Goliat field section}  \label{sec:results}

The Kobbe Formation of the Goliat field is the region/section under consideration for realistic inversion and estimation. The Kobbe Formation is deposited in Middle Triassic age and represents a prograding deltaic system with mouth bars and tidally influenced lobes. In its lower section, the system shifts into a more proximal, heterogeneous fluvial setting. A total of nine stratigraphic zones have been identified: Kobbe 1 to Kobbe 9 from base to top. The Kobbe reservoir is split into Lower Kobbe (Kobbe 1-7) and Upper Kobbe (Kobbe 8 and 9) \citep{larsen2015extra}.

The Upper Kobbe formation represents a prograding delta front environment. Kobbe 9 consists predominantly of fine-grained sandstones, interpreted to be deltaic mouth bar deposits, interbedded with coarse grained levels interpreted to be fluvial channels \citep{larsen2015extra}. The deltaic system is prograding towards the west, and an increase in shale content is expected towards the west (prodelta fining deposits) and the east (flood plain shales). From the geological model, well A is entering the proximal part of the delta in an area where it is expected to find high amount of fluvial channel deposits \citep{larsen2015extra}.

The prior geo-models of three layers and four layers are selected after checking for non-uniqueness and multi-modality for the given section of the Goliat field. In the next step, the inversion is performed on the real data of the Goliat field section and selected prior geo-models. An ensemble of 1000 members with 8 numbers of iterations have been used in all cases of real-time inversion of Goliat field section. The inversions are performed in real-time using the combined framework of FlexIES and DNN, which takes around 60 to 80 seconds for 9000 functions evaluations. Furthermore, the inversions are also performed while neglecting model-error using ESMDA, which can be useful to compare the uncertainties associated in the presence of the model error with FlexIES.

\subsection{Considering prior geo-model of three layers}
\label{sec:case1}

\begin{figure}
\begin{center}

 \hspace{-0.5in}
    \begin{subfigure}[normal]{0.4\textwidth}
	\includegraphics[scale=0.5]{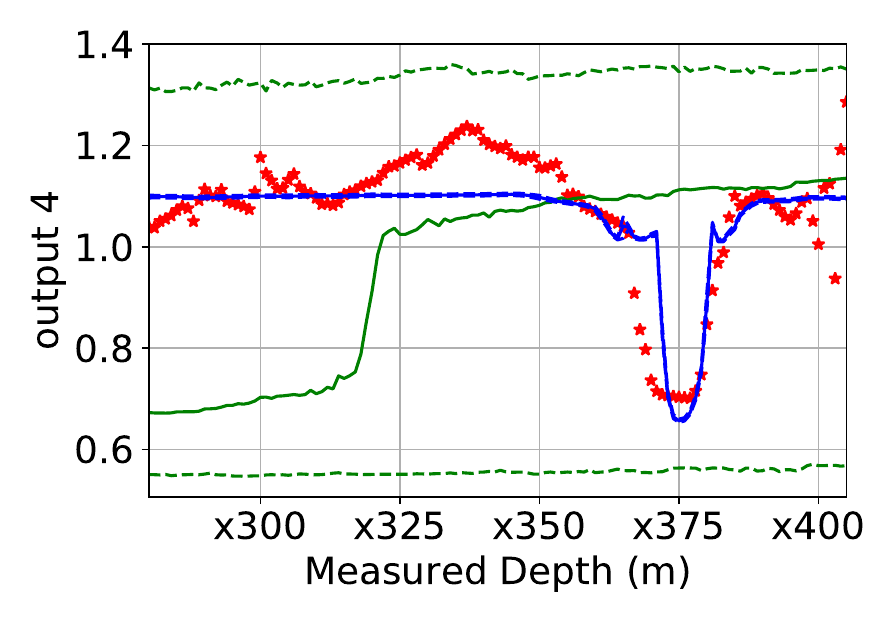}	
	\end{subfigure}
 \hspace{0.5in}
	\begin{subfigure}[normal]{0.4\textwidth}
	\includegraphics[scale=0.5]{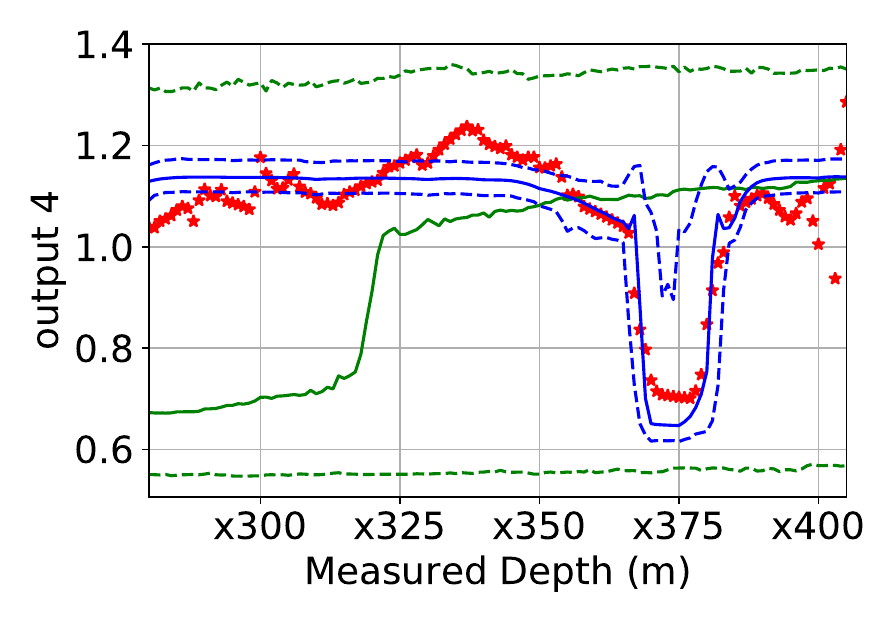}	
	\end{subfigure}

 \hspace{-0.5in}
    \begin{subfigure}[normal]{0.4\textwidth}
	\includegraphics[scale=0.5]{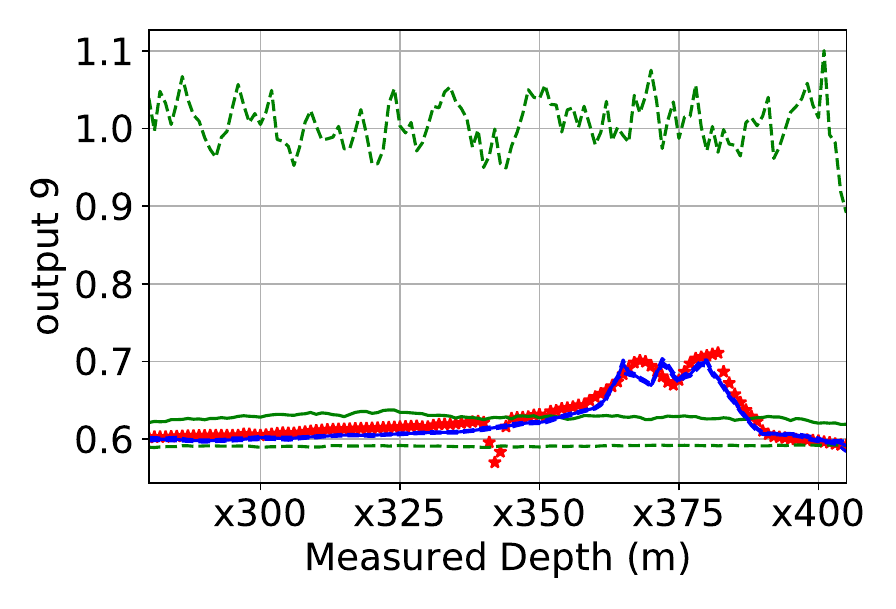}	
	\end{subfigure}
 \hspace{0.5in}
	\begin{subfigure}[normal]{0.4\textwidth}
	\includegraphics[scale=0.5]{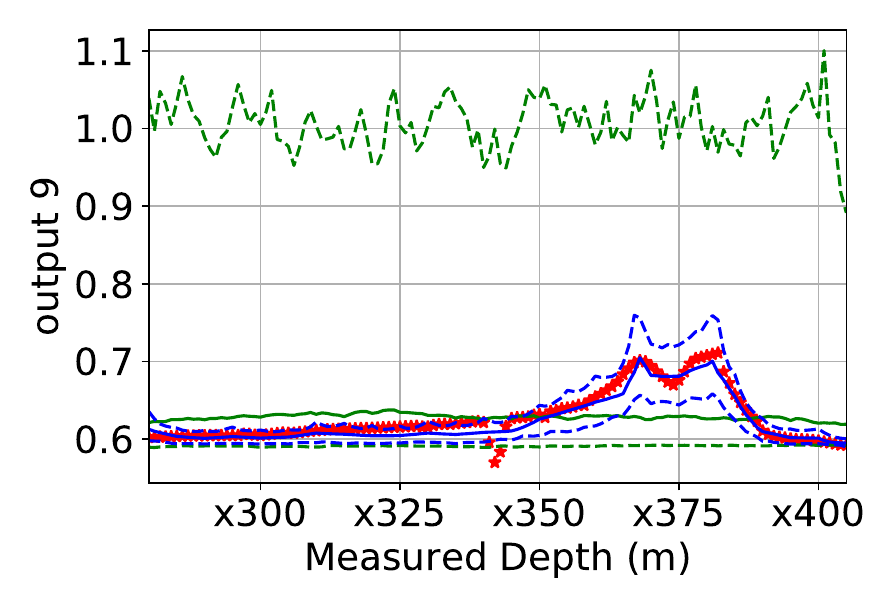}	
	\end{subfigure}

 \hspace{-0.5in}
    \begin{subfigure}[normal]{0.4\textwidth}
	\includegraphics[scale=0.5]{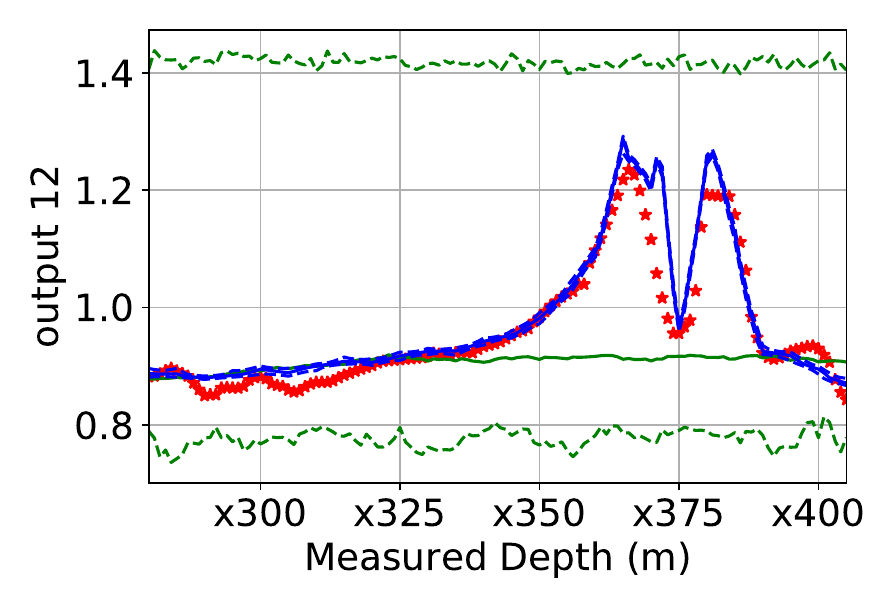}	
	\end{subfigure}
 \hspace{0.5in}
	\begin{subfigure}[normal]{0.4\textwidth}
	\includegraphics[scale=0.5]{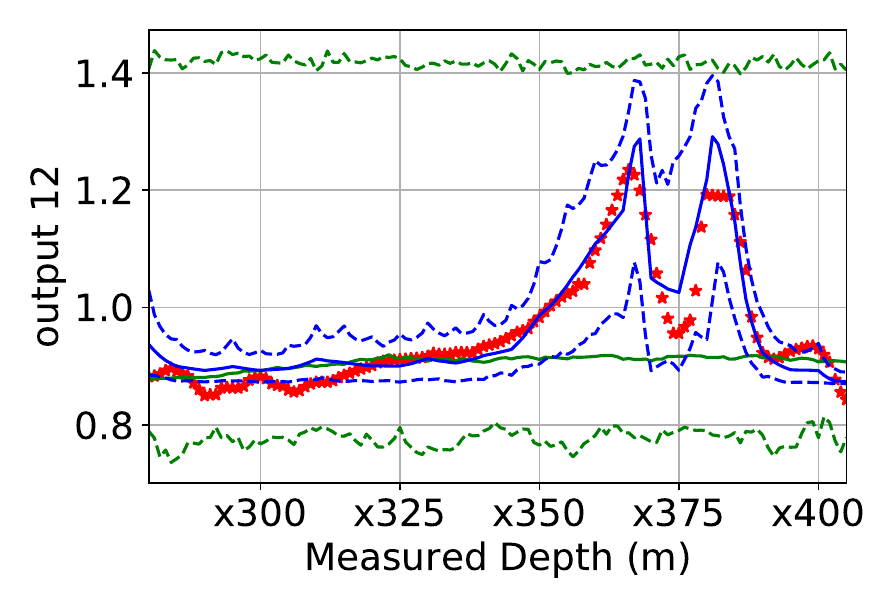}	
	\end{subfigure}

 \hspace{0.4in}
		 ESMDA (neglecting model-error)
 \hspace{0.7in}
 		 FlexIES (accounting for model-error)
 \end{center}

\caption{Inversion results of EM outputs considering prior geo-model of three layers. Green and blue lines show ensemble approximation of the prior and posterior distribution respectively and red stars show observed EM measurements. Solid green and solid blue lines show 50th percentile ($p50$) and dashed green and dashed blue lines show $98\%$ confidence interval respectively of the prior and posterior distribution respectively. First and second column of the sub-figures show inversion results obtained from ESMDA and FlexIES respectively.}
\label{post_outputs_2_1}
\end{figure}

\begin{figure}
\begin{center}

 \hspace{-0.5in}
    \begin{subfigure}[normal]{0.4\textwidth}
	\includegraphics[scale=0.5]{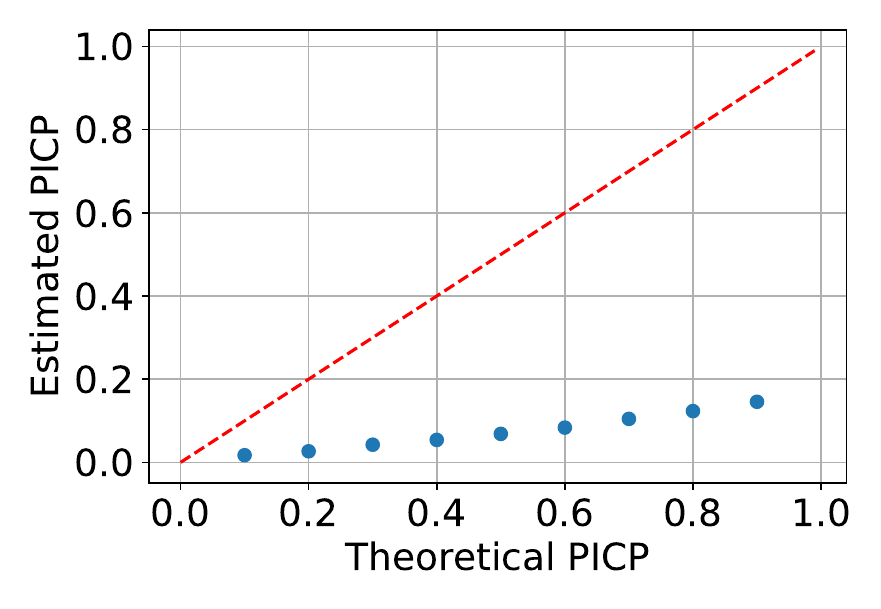}	
	\end{subfigure}
 \hspace{0.5in}
	\begin{subfigure}[normal]{0.4\textwidth}
	\includegraphics[scale=0.5]{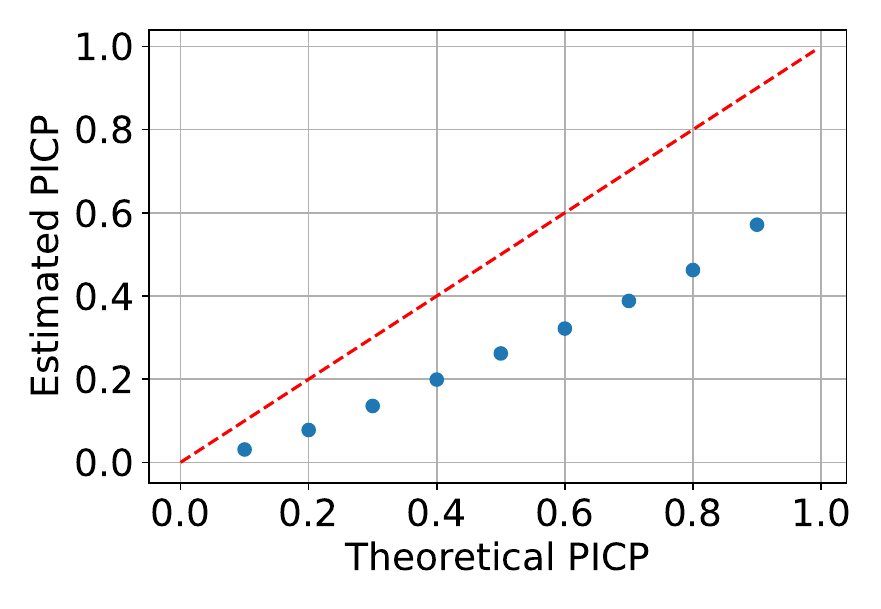}	
	\end{subfigure}

 \hspace{0.4in}
		 ESMDA (neglecting model-error)
 \hspace{0.7in}
 		 FlexIES (accounting for model-error)		 
 		 
\end{center} 
 
\caption{PICP considering prior geo-model of three layers.}
\label{picp_1}
\end{figure}

In this case, the prior geo-model of three layers i.e.~shale-sand-shale sequence is considered for the realistic inversion of the Goliat field section. Figure \ref{post_outputs_2_1} shows the inversion results of the EM outputs considering the prior geo-model of the three layers while neglecting and accounting for model error. The EM outputs show reasonable match of the patterns of the EM observations in the absence of the model error, however some points are not completely matched due to the negligence of the model error. The uncertainties associated with model error are captured using FlexIES by covering unmatched points of the EM outputs as shown in the right panel of the Figure \ref{post_outputs_2_1}. The quality of the inversion of the EM measurements are assessed using prediction interval coverage probability (PICP) and mean continuous ranked probability score (CRPS). The complete mathematical descriptions of PICP and CRPS are shown in the Appendix \ref{sec:AppendixB}. The PICP is used to assess the associated uncertainties in the EM measurements related to the model error and measurement error \citep{xu2015data}. PICP values close to the $45^o$ line (dashed red line) indicate a perfect posterior distribution. Figure \ref{picp_1} shows that the estimated PICP with respect to the theoretical PICP are very far from the diagonal line while neglecting model error using ESMDA. However, the estimated PICP values are improved while accounting for model error using FlexIES.

\begin{figure}
\begin{center}

 \hspace{-0.5in}
    \begin{subfigure}[normal]{0.4\textwidth}
	\includegraphics[scale=0.5]{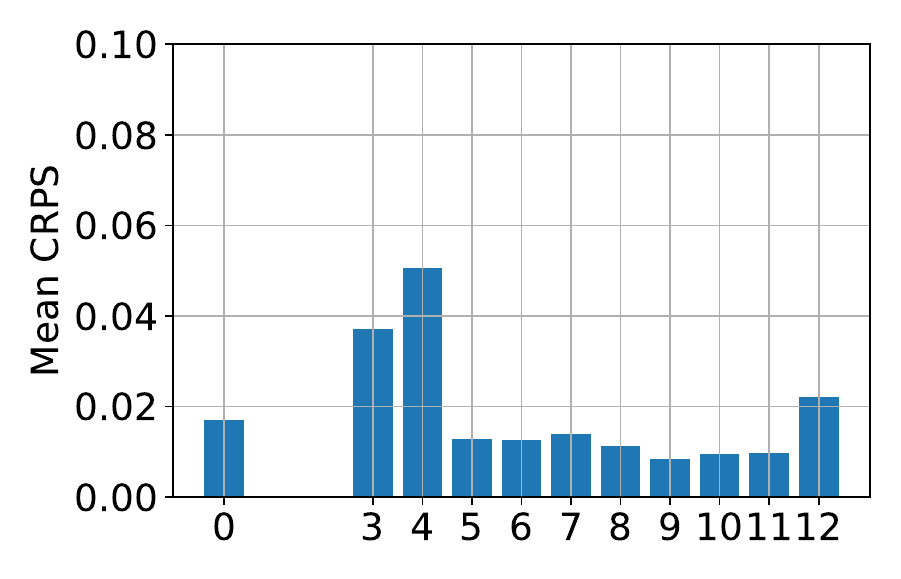}	
	\end{subfigure}
 \hspace{0.5in}
	\begin{subfigure}[normal]{0.4\textwidth}
	\includegraphics[scale=0.5]{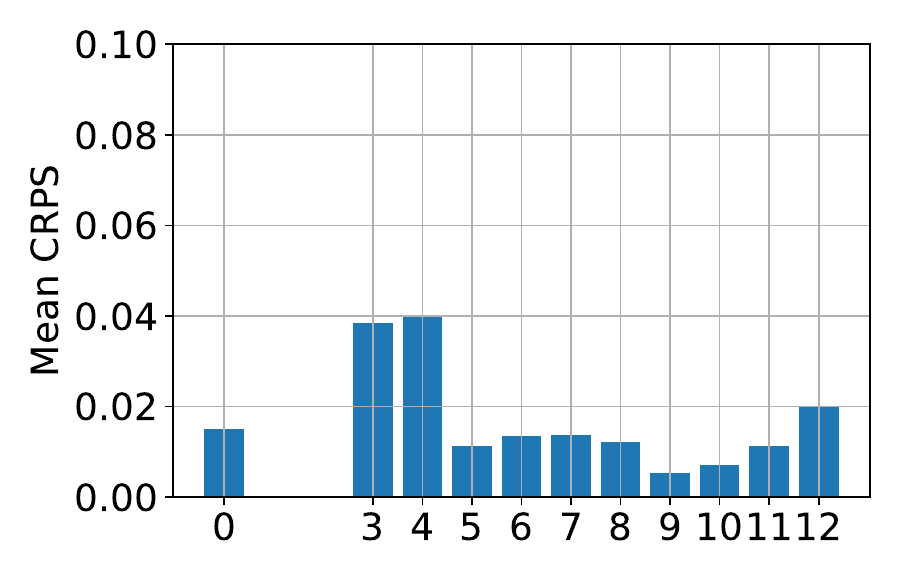}	
	\end{subfigure}

 \hspace{0.4in}
		 ESMDA (neglecting model-error)
 \hspace{0.7in}
 		 FlexIES (accounting for model-error)		 
 		 
\end{center} 
 
\caption{Mean CRPS considering prior geo-model of three layers. X-axis shows the log numbers.}
\label{crps_1}
\end{figure}

Figure \ref{crps_1} shows the mean CRPS of the posterior ensemble of the outputs of the EM measurements. We observe slight improvement in the mean CRPS values of the EM outputs while accounting for model error. This is due to the fact that in this case we did not include observed EM measurements of logs 1 and 2 i.e.~(RT-CRES-RPCEHX; RT-CRES-RACEHX) that exhibit large model errors.

\begin{figure}
\begin{center}

 \hspace{-0.7in}
    \begin{subfigure}[normal]{0.4\textwidth}
	\includegraphics[scale=0.5]{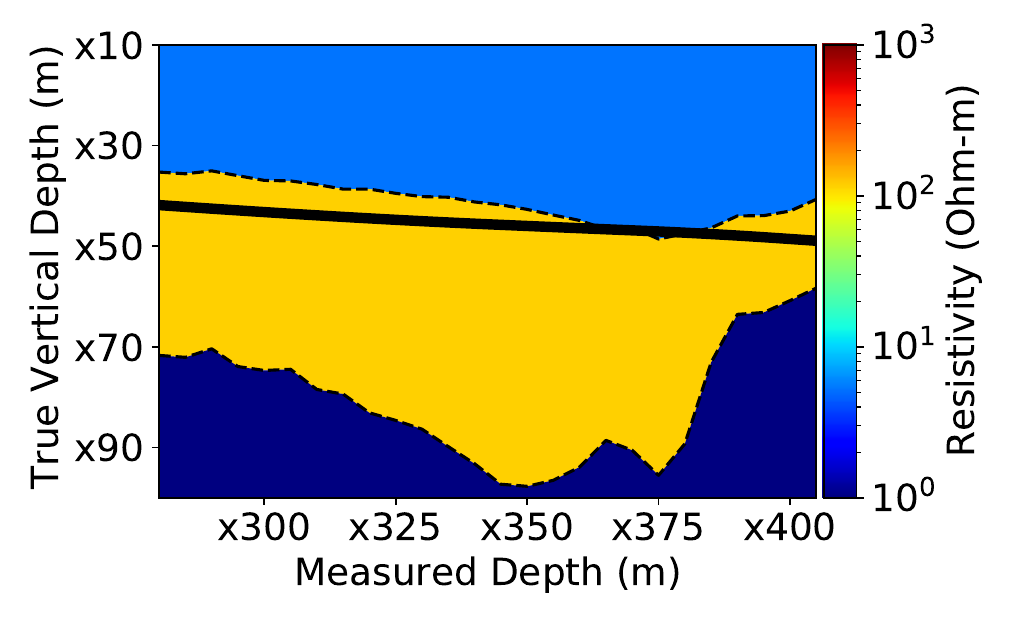}	
	\end{subfigure}
 \hspace{0.45in}
	\begin{subfigure}[normal]{0.4\textwidth}
	\includegraphics[scale=0.5]{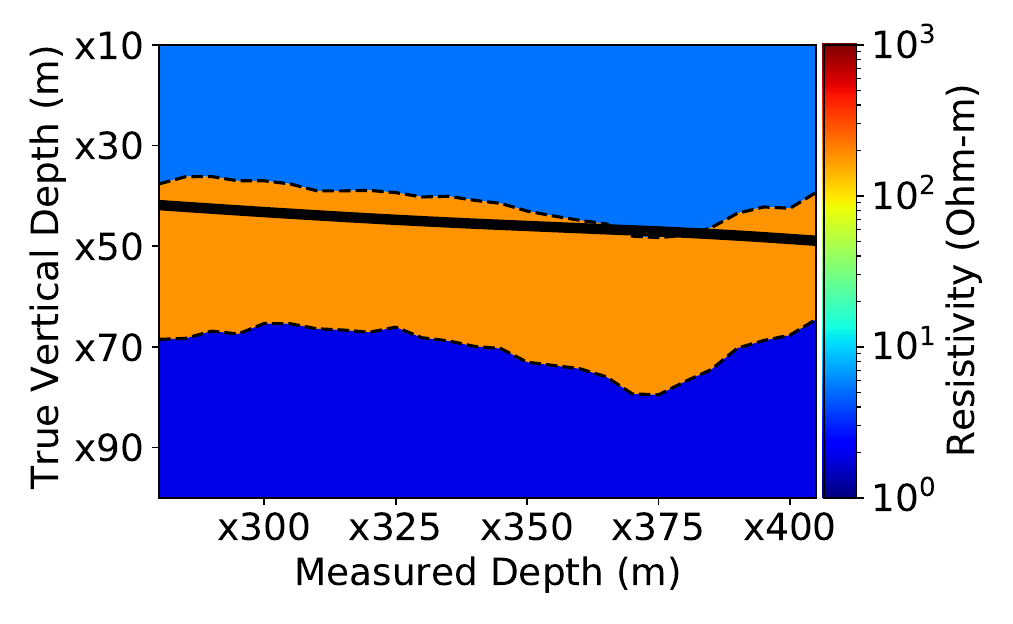}	
	\end{subfigure}

 \hspace{0.4in}
		 ESMDA (neglecting model-error)
 \hspace{0.7in}
 		 FlexIES (accounting for model-error)
 \end{center}

\caption{Estimation results of the geo-layers profile considering prior geo-model of three layers.}
\label{est_2a}
\end{figure}

\begin{figure}
\begin{center}

 \hspace{-0.5in}
    \begin{subfigure}[normal]{0.4\textwidth}
	\includegraphics[scale=0.5]{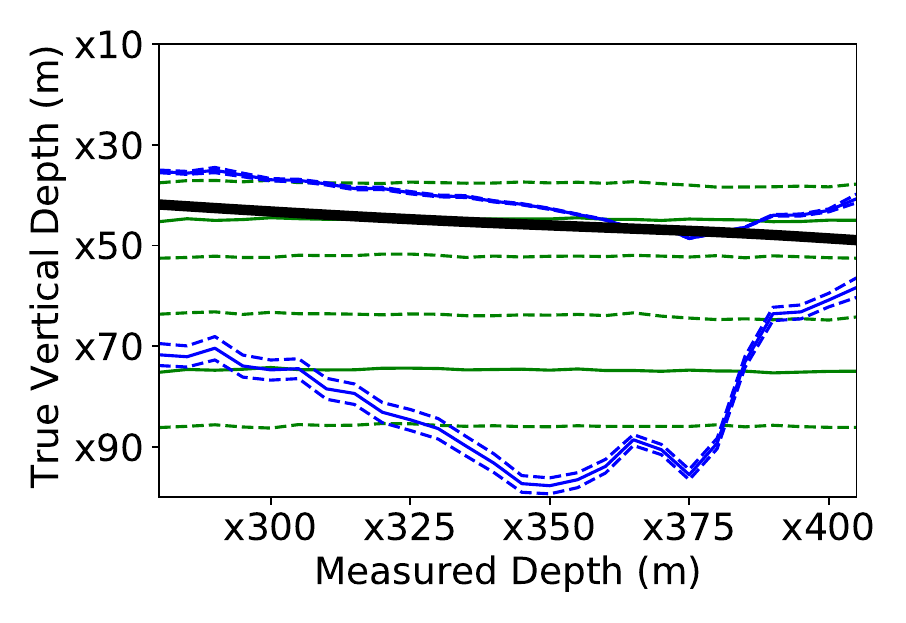}	
	\end{subfigure}
 \hspace{0.5in}
	\begin{subfigure}[normal]{0.4\textwidth}
	\includegraphics[scale=0.5]{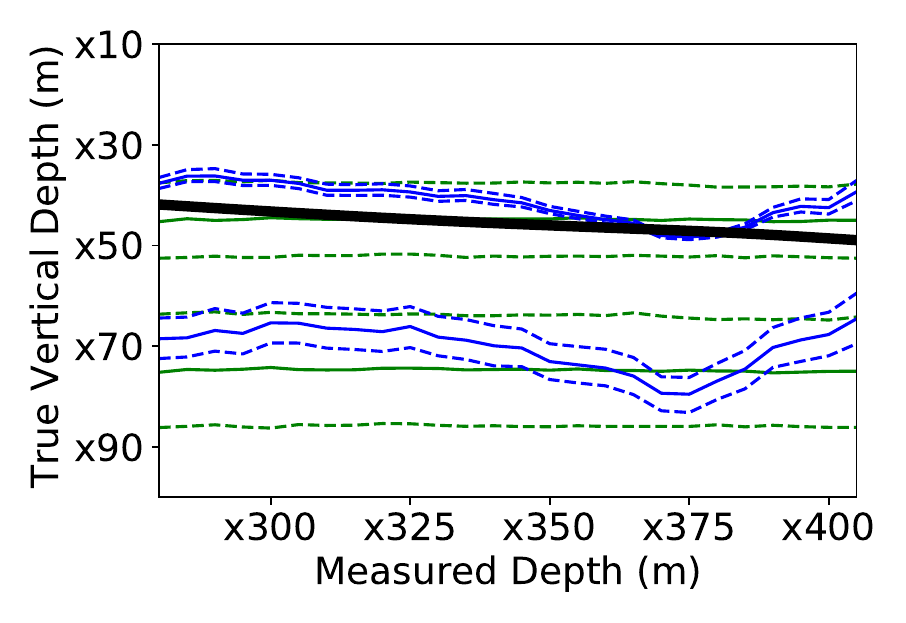}	
	\end{subfigure}

 \hspace{0.4in}
		 ESMDA (neglecting model-error)
 \hspace{0.7in}
 		 FlexIES (accounting for model-error)
 \end{center}

\caption{Prior and posterior distribution of the geo-layers profile considering three layers. Green and blue lines show ensemble approximation of the prior and posterior distribution respectively. Solid green and solid blue lines show 50th percentile ($p50$) and dashed green and dashed blue lines show $80\%$ confidence interval of the prior and posterior distribution respectively.}
\label{est_2b}
\end{figure}

Figure \ref{est_2a} shows the estimation results of the geo-layers profile of the Goliat field section. The estimated profiles of the bottom boundaries and resitivities are different while neglecting and accounting for model error. Furthermore, the resistivities are also different using ESMDA and FlexIES. This effect can be attributed to the negligence and accounting for model error during inversion. However, the  estimated results from FlexIES are closer to the proprietary inversion from the tool vendor as shown in Figure \ref{prop_inversion}. Figure \ref{est_2b} shows the ensemble approximation of the prior and posterior distribution of the geo-layers profile. Low confidence intervals are observed in geo-layers profile while neglecting model error, that indicates the underestimation of the uncertainties. The quantification of the uncertainties are improved in geo-layer profiles while accounting for model error using FlexIES. The bottom layer boundaries show higher uncertainties as compared to the upper layer boundaries because the well trajectory is far from the bottom layer boundary.  

\subsection{Considering prior geo-model of four layers}\label{sec:case2}

\begin{figure}
\begin{center}

 \hspace{-0.5in}
    \begin{subfigure}[normal]{0.4\textwidth}
	\includegraphics[scale=0.5]{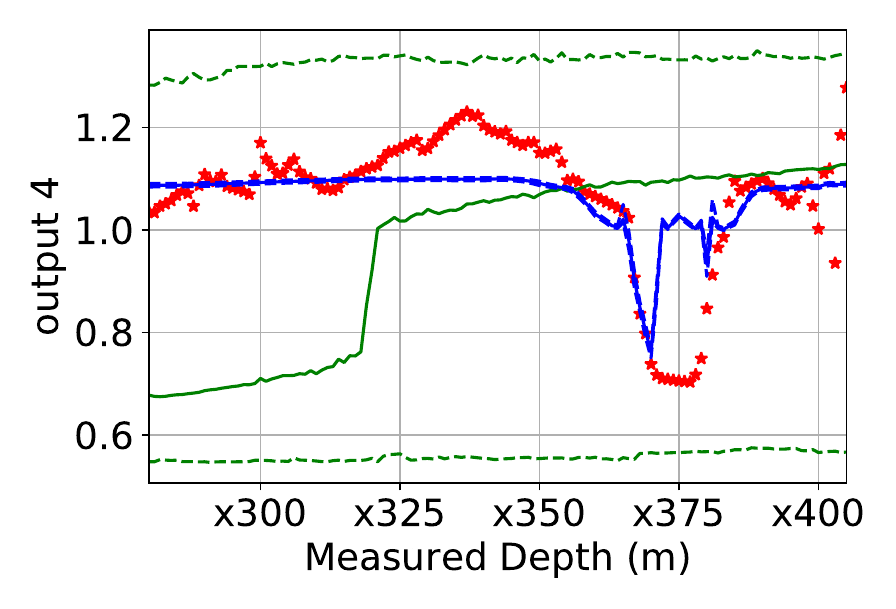}	
	\end{subfigure}
 \hspace{0.5in}
	\begin{subfigure}[normal]{0.4\textwidth}
	\includegraphics[scale=0.5]{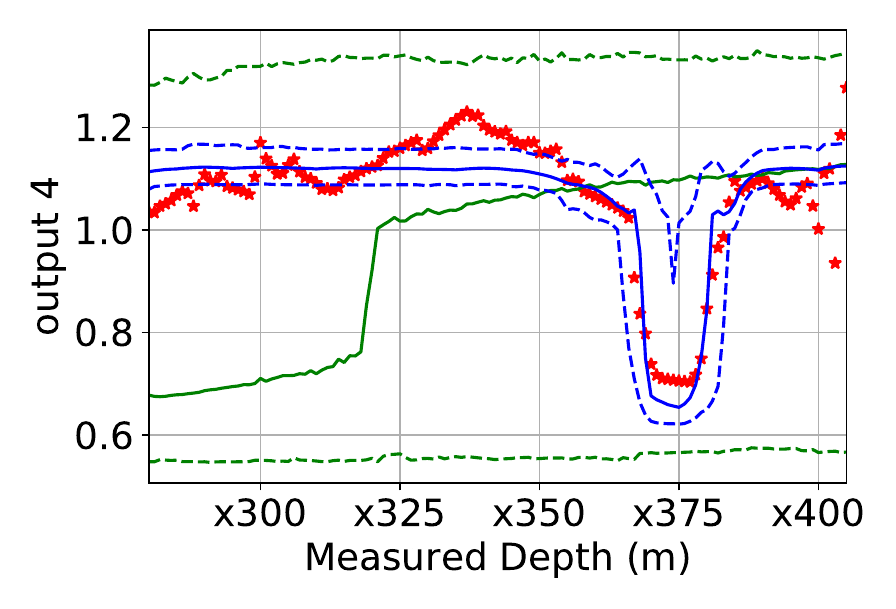}	
	\end{subfigure}

 \hspace{-0.5in}
    \begin{subfigure}[normal]{0.4\textwidth}
	\includegraphics[scale=0.5]{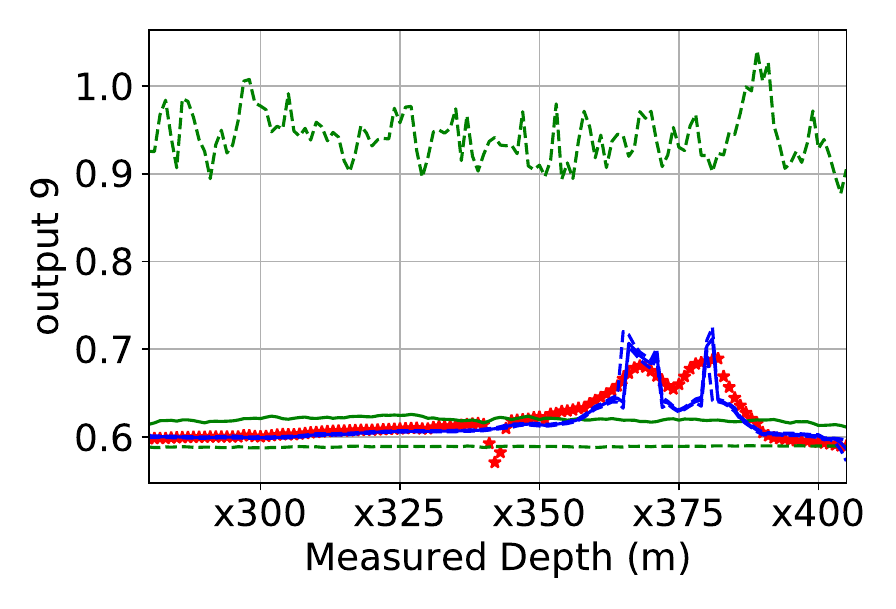}	
	\end{subfigure}
 \hspace{0.5in}
	\begin{subfigure}[normal]{0.4\textwidth}
	\includegraphics[scale=0.5]{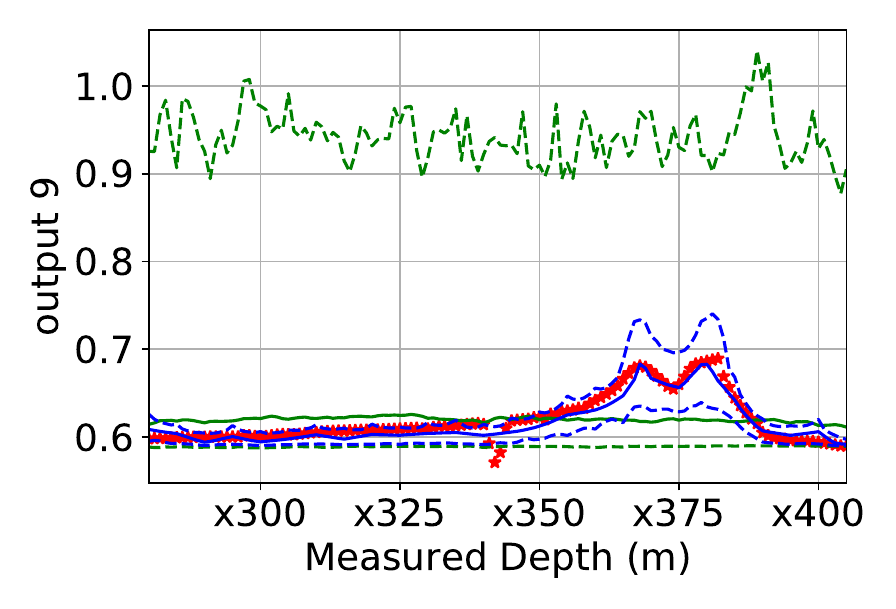}	
	\end{subfigure}

 \hspace{-0.5in}
    \begin{subfigure}[normal]{0.4\textwidth}
	\includegraphics[scale=0.5]{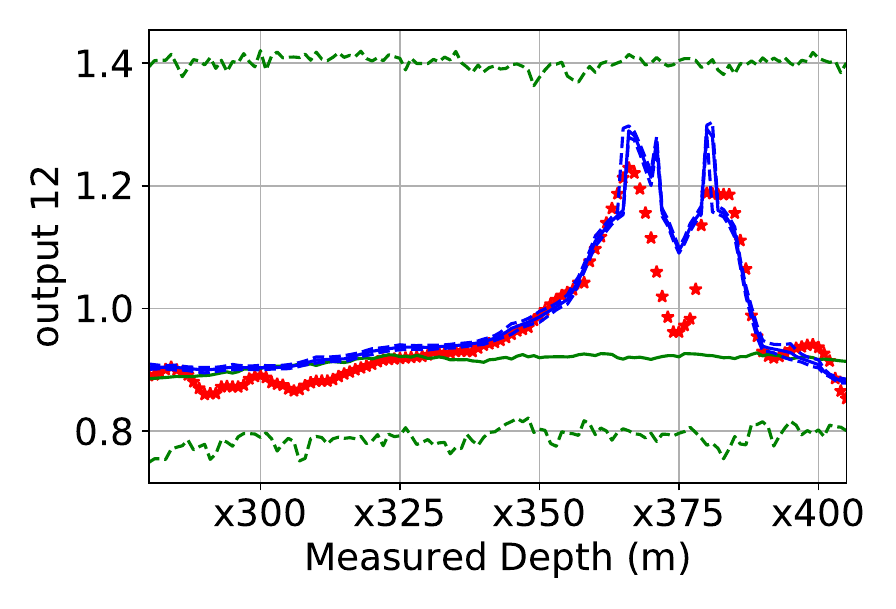}	
	\end{subfigure}
 \hspace{0.5in}
	\begin{subfigure}[normal]{0.4\textwidth}
	\includegraphics[scale=0.5]{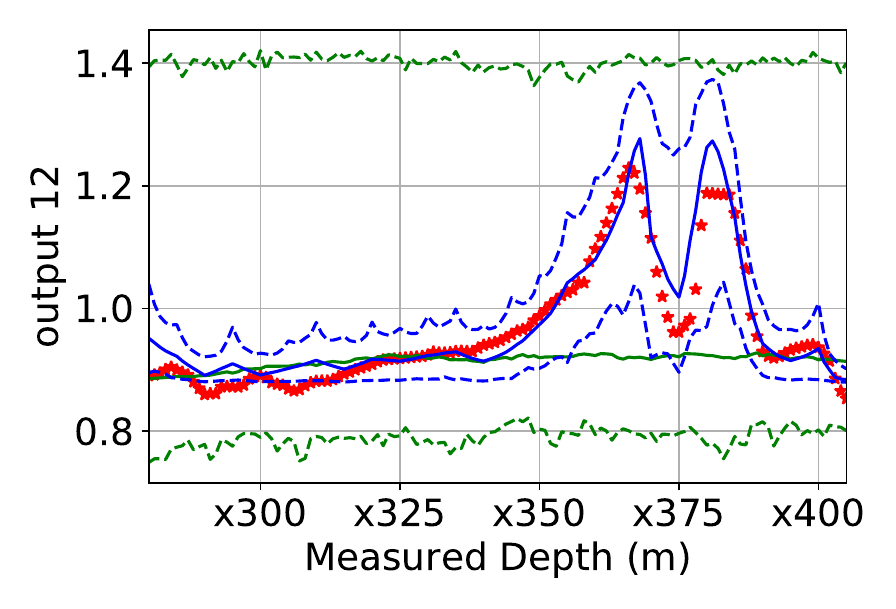}	
	\end{subfigure}

 \hspace{0.4in}
		 ESMDA (neglecting model-error)
 \hspace{0.7in}
 		 FlexIES (accounting for model-error)
 \end{center}

\caption{Inversion results of EM outputs considering prior geo-model of four layers. Green and blue lines show ensemble approximation of the prior and posterior distribution respectively and red stars show observed EM measurements. Solid green and solid blue lines show 50th percentile ($p50$) and dashed green and dashed blue lines show $98\%$ confidence interval respectively of the prior and posterior distribution respectively. First and second column of the sub-figures show inversion results obtained from ESMDA and FlexIES respectively.}
\label{post_outputs_3_1}
\end{figure}

In this case, the prior geo-model of four layers is considered for the realistic inversion of the Goliat field section. This case allows us to explore the further improvement in the real-time inversion of the Goliat field section as compared to the three layers geo-model.
The inversion and data matching results of the EM outputs are similar to the previous case while neglecting and accounting for model error as shown in Figure \ref{post_outputs_3_1}. FlexIES covers the unmatched points of the EM outputs that indicates the algorithm tries to cover the uncertainties associated with the model error. 

\begin{figure}
\begin{center}

 \hspace{-0.5in}
    \begin{subfigure}[normal]{0.4\textwidth}
	\includegraphics[scale=0.5]{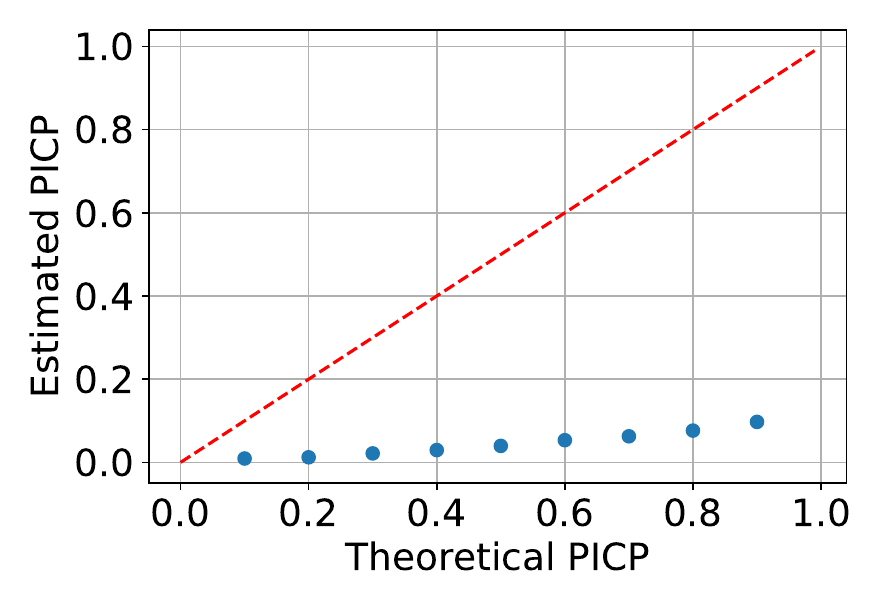}	
	\end{subfigure}
 \hspace{0.5in}
	\begin{subfigure}[normal]{0.4\textwidth}
	\includegraphics[scale=0.5]{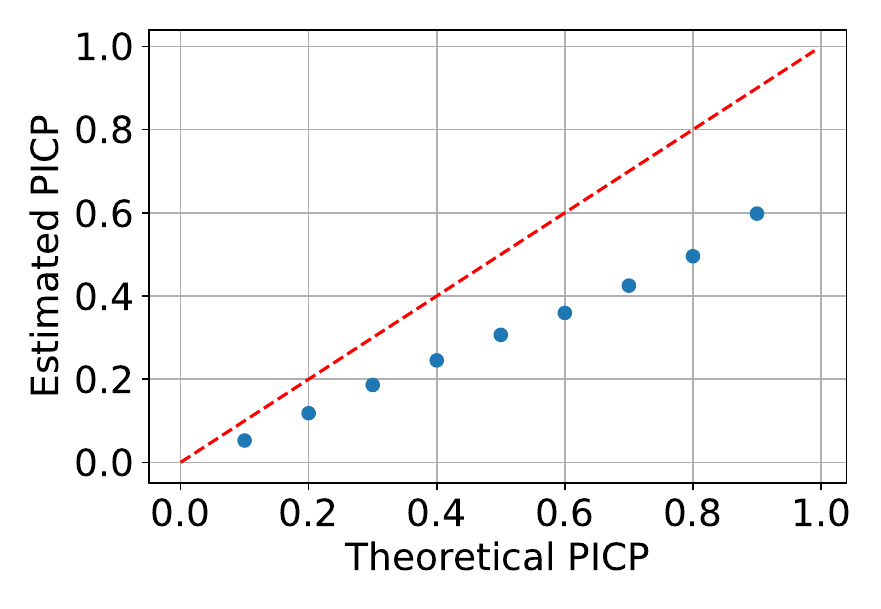}	
	\end{subfigure}

 \hspace{0.4in}
		 ESMDA (neglecting model-error)
 \hspace{0.7in}
 		 FlexIES (accounting for model-error)		 
 		 
\end{center} 
 
\caption{PICP considering prior geo-model of four layers.}
\label{picp_2}
\end{figure}

\begin{figure}
\begin{center}

 \hspace{-0.5in}
    \begin{subfigure}[normal]{0.4\textwidth}
	\includegraphics[scale=0.5]{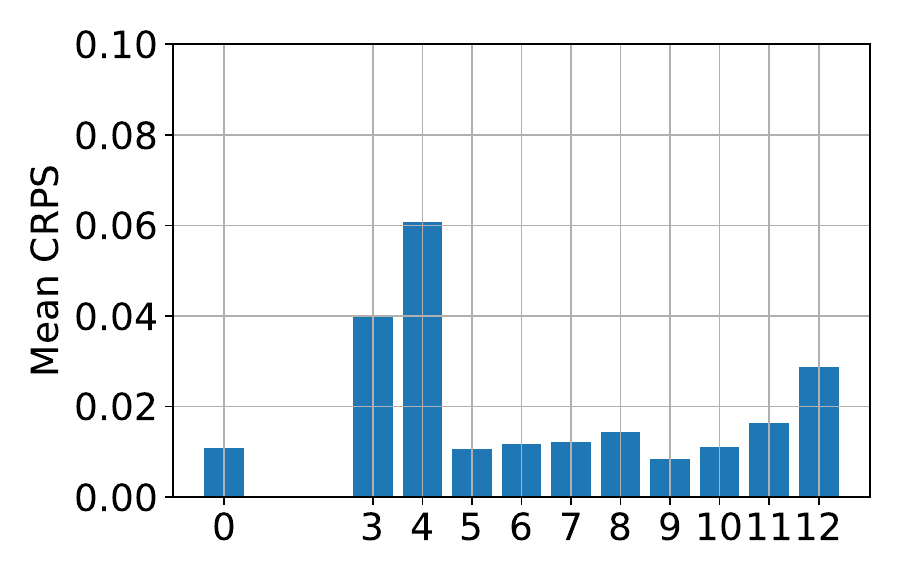}	
	\end{subfigure}
 \hspace{0.5in}
	\begin{subfigure}[normal]{0.4\textwidth}
	\includegraphics[scale=0.5]{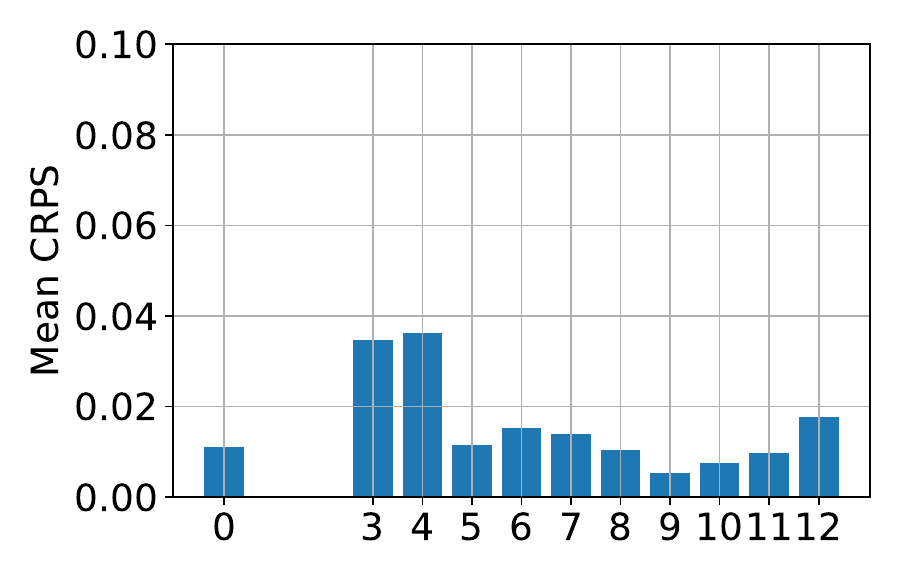}	
	\end{subfigure}

 \hspace{0.4in}
		 ESMDA (neglecting model-error)
 \hspace{0.7in}
 		 FlexIES (accounting for model-error)		 
 		 
\end{center} 
 
\caption{Mean CRPS considering prior geo-model of four layers. X-axis shows the log numbers.}
\label{crps_2}
\end{figure}

Similar to the previous case, the quality of the inversion of the EM measurements is also assessed using PICP and mean CRPS. We observed similar improvement in PICP and mean CRPS values while accounting for model error as shown in Figures \ref{picp_2} and \ref{crps_2}. However PICP and mean CRPS values are slightly improved for four layer case as compared to the previous three layer case using FlexIES. 

\begin{figure}
\begin{center}

 \hspace{-0.7in}
    \begin{subfigure}[normal]{0.4\textwidth}
	\includegraphics[scale=0.5]{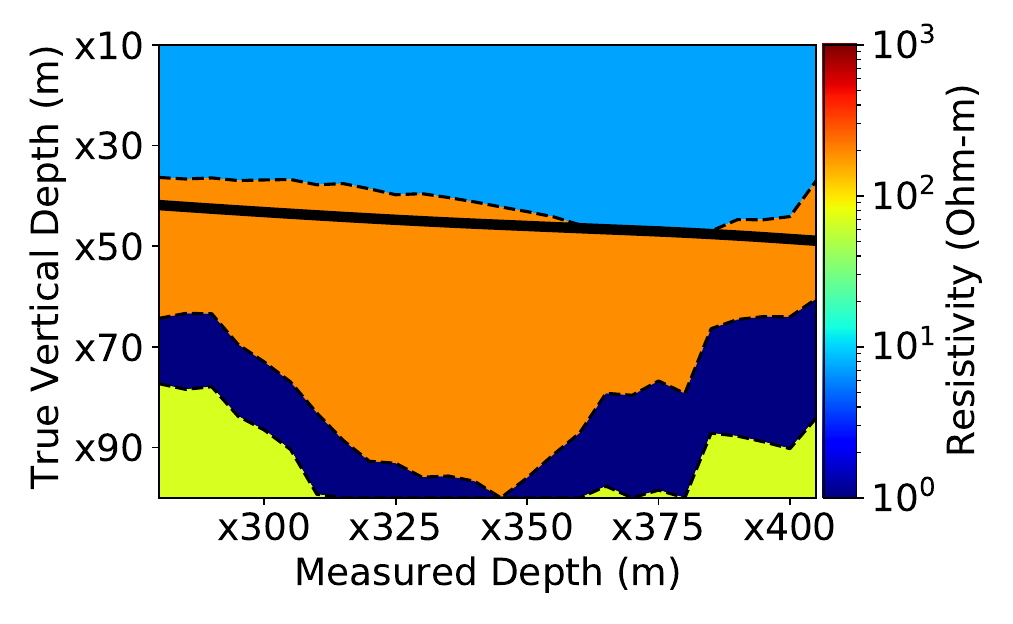}	
	\end{subfigure}
 \hspace{0.45in}
	\begin{subfigure}[normal]{0.4\textwidth}
	\includegraphics[scale=0.5]{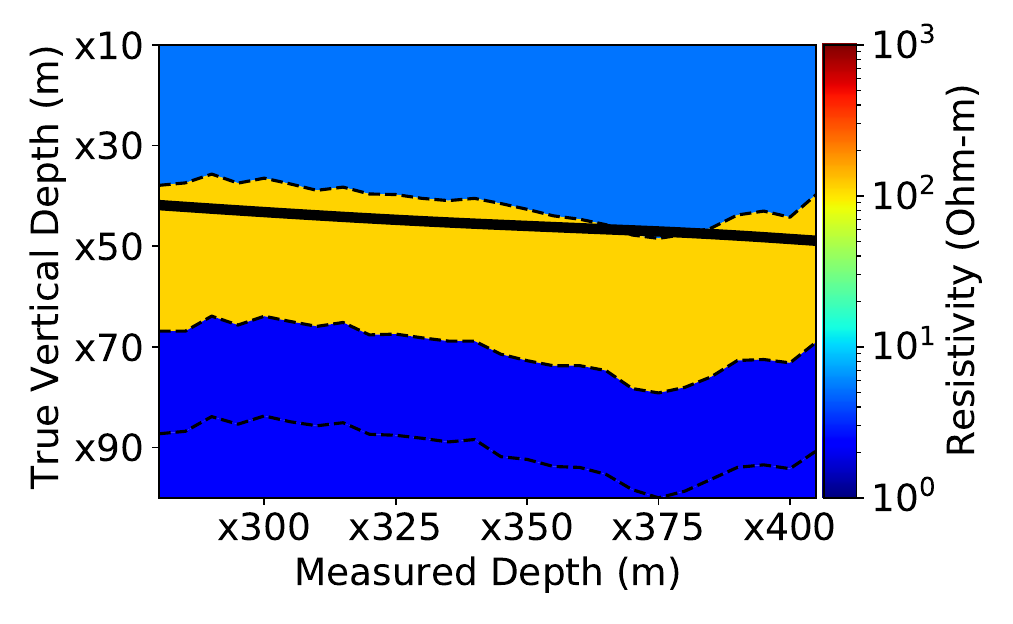}	
	\end{subfigure}

 \hspace{0.4in}
		 ESMDA (neglecting model-error)
 \hspace{0.7in}
 		 FlexIES (accounting for model-error)
 \end{center}

\caption{Estimation results of the geo-layers profile considering prior geo-model of four layers.
}
\label{est_3a}
\end{figure}

Figure \ref{est_3a} shows the estimation results of the geo-layers profile of the Goliat field section by considering the prior geo-model of four layers. Similar to the previous case, the estimated profiles of the bottom boundaries and resitivities are different while neglecting and accounting for model error along with resistivities. Furthermore, the third and fourth layers also show the difference in resistivities while neglecting model error. However, the  estimated results from FlexIES are similar to the previous case thus showing no contrast in the resistivities of the third and fourth layers.  

\begin{figure}
\begin{center}

 \hspace{-0.5in}
    \begin{subfigure}[normal]{0.4\textwidth}
	\includegraphics[scale=0.5]{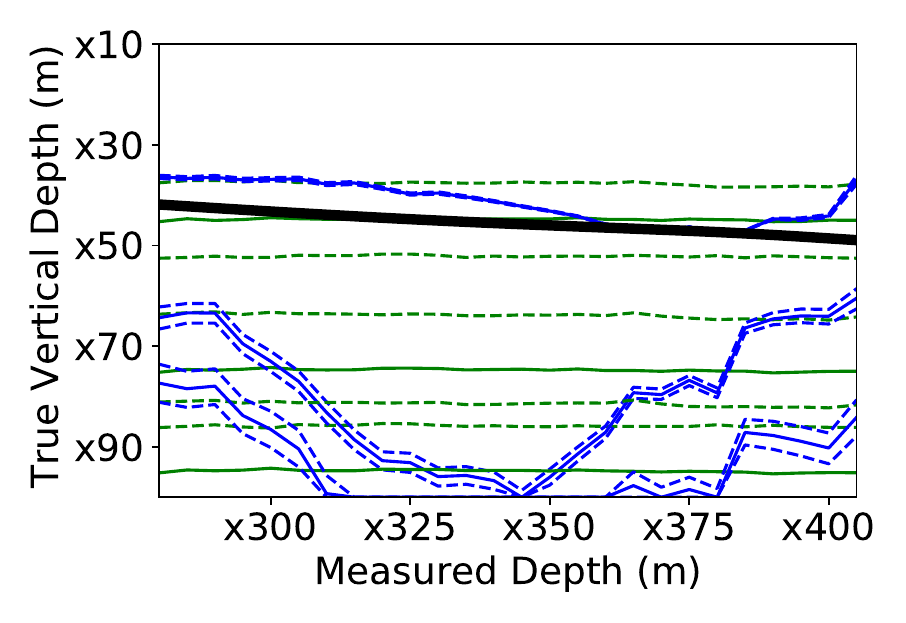}	
	\end{subfigure}
 \hspace{0.5in}
	\begin{subfigure}[normal]{0.4\textwidth}
	\includegraphics[scale=0.5]{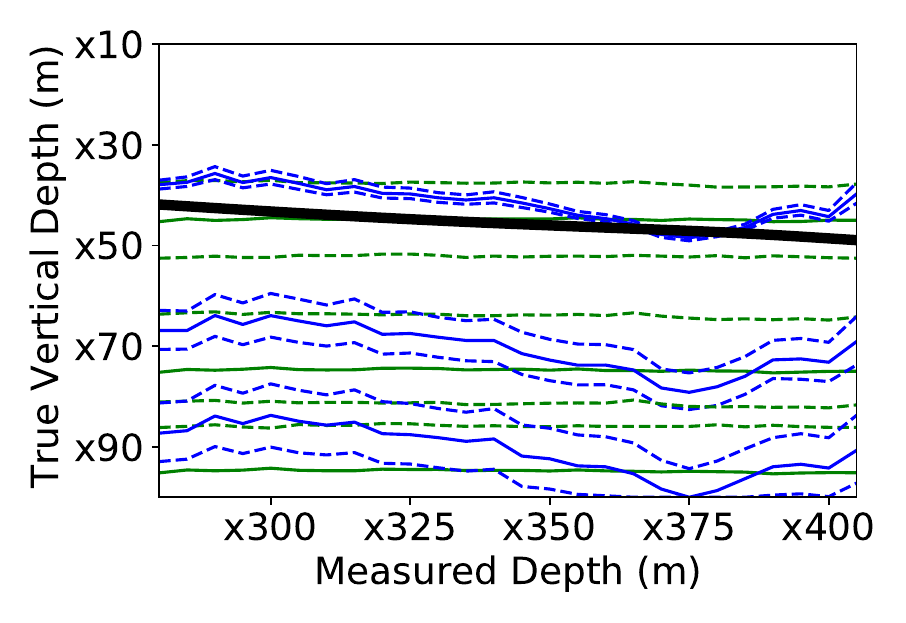}	
	\end{subfigure}

\hspace{0.4in}
		 ESMDA (neglecting model-error)
 \hspace{0.7in}
 		 FlexIES (accounting for model-error)
 \end{center}

\caption{Prior and posterior distribution of the geo-layers profile considering four layers. Green and blue lines show ensemble approximation of the prior and posterior distribution respectively. Solid green and solid blue lines show 50th percentile ($p50$) and dashed green and dashed blue lines show $80\%$ confidence interval of the prior and posterior distribution respectively.}
\label{est_3b}
\end{figure}

Figure \ref{est_3b} shows the ensemble approximation of the prior and posterior distribution of the geo-layers profile. Similar to the previous case low confidence intervals are observed in geo-layers profile while neglecting model error, that indicates the underestimation of the uncertainties. The quantification of the uncertainties are improved in geo-layer profiles while accounting for model error using FlexIES.

\subsection{Assimilating data with large model errors}

\begin{figure}
\begin{center}

 \hspace{-0.5in}
    \begin{subfigure}[normal]{0.4\textwidth}
	\includegraphics[scale=0.5]{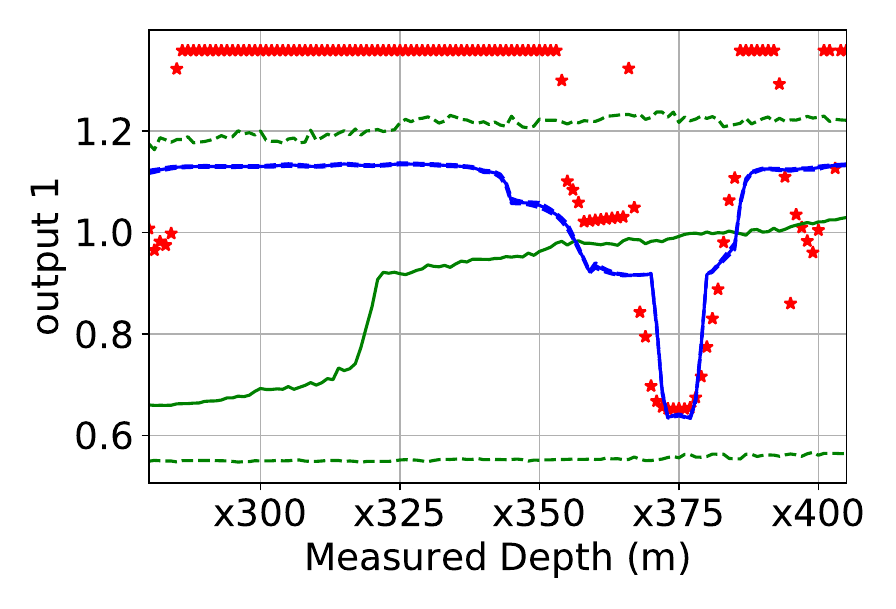}	
	\end{subfigure}
 \hspace{0.5in}
	\begin{subfigure}[normal]{0.4\textwidth}
	\includegraphics[scale=0.5]{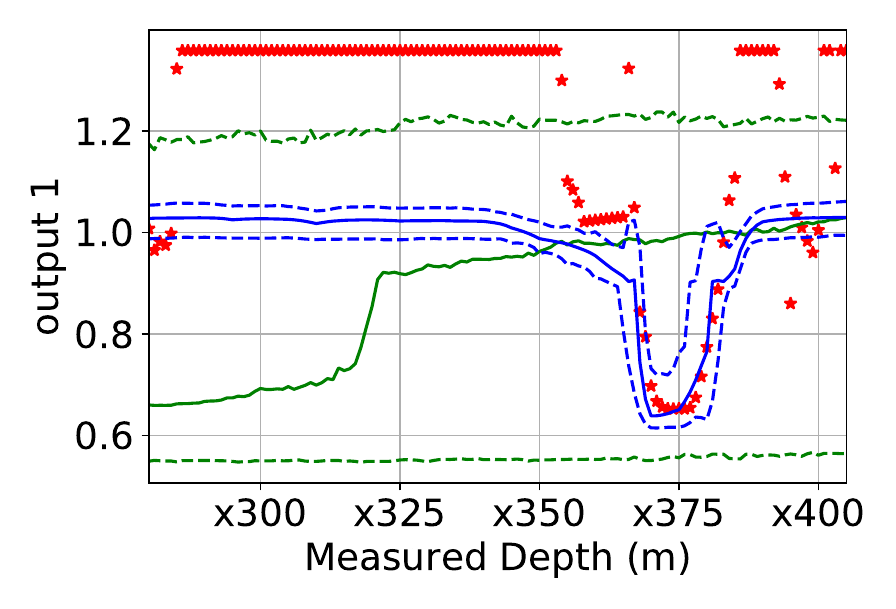}	
	\end{subfigure}

 \hspace{-0.5in}
    \begin{subfigure}[normal]{0.4\textwidth}
	\includegraphics[scale=0.5]{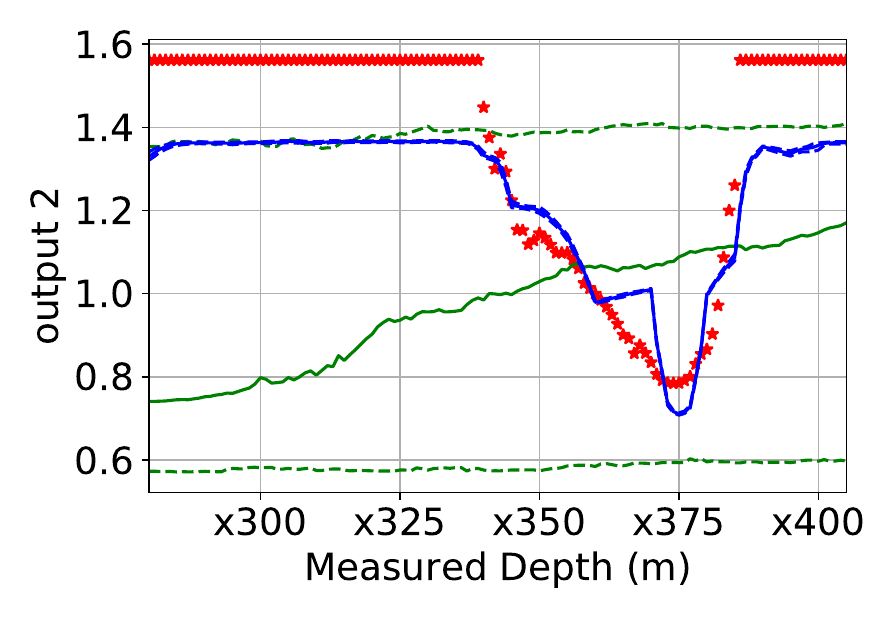}	
	\end{subfigure}
 \hspace{0.5in}
	\begin{subfigure}[normal]{0.4\textwidth}
	\includegraphics[scale=0.5]{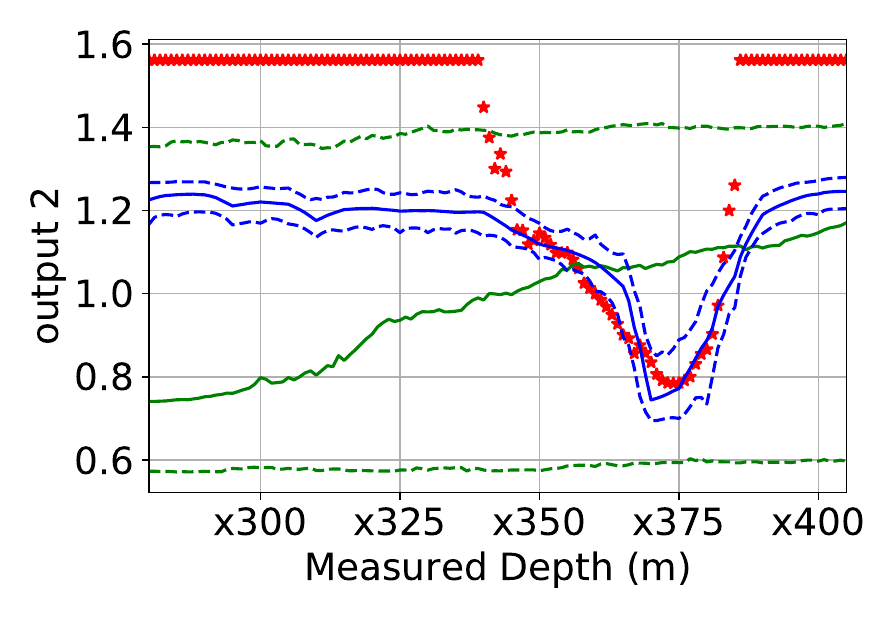}	
	\end{subfigure}
	
 \hspace{0.4in}
		 ESMDA (neglecting model-error)
 \hspace{0.7in}
 		 FlexIES (accounting for model-error)
 \end{center}

\caption{Prior and posterior distribution of the geo-layers profile considering three layers. In the first row of the sub-figures, green and blue lines show ensemble approximation of the prior and posterior distribution respectively. Solid green and solid blue lines show 50th percentile ($p50$) and dashed green and dashed blue lines show $98\%$ confidence interval respectively of the prior and posterior distribution respectively.}
\label{post_output_4_2}
\end{figure}

For the case of deep EM measurements, the shallowest (2 MHz, output 1 / 400 kHz, output 2) logs are sensitive to the local variations of the earth model. These variations are not important for geosteering decisions, but might influence interpretation. 
 Figure \ref{post_output_4_2} shows the ensemble approximation of the prior and posterior distribution of the erroneous outputs 1 and 2. We note that the deep non-directional attenuation outputs 1 and 2  are outside the prior range when drilling in the second layer of the geo-model. Furthermore, we note that the prior does not cover the data completely which can be attributed to the large model error for outputs 1 and 2 as shown in Figure \ref{fig:realLogsDeep}. Moreover, Figure \ref{post_output_4_2} shows no reasonable match of the data for the outputs 1 and 2. 

\begin{figure}
\begin{center}

 \hspace{-0.7in}
    \begin{subfigure}[normal]{0.4\textwidth}
	\includegraphics[scale=0.5]{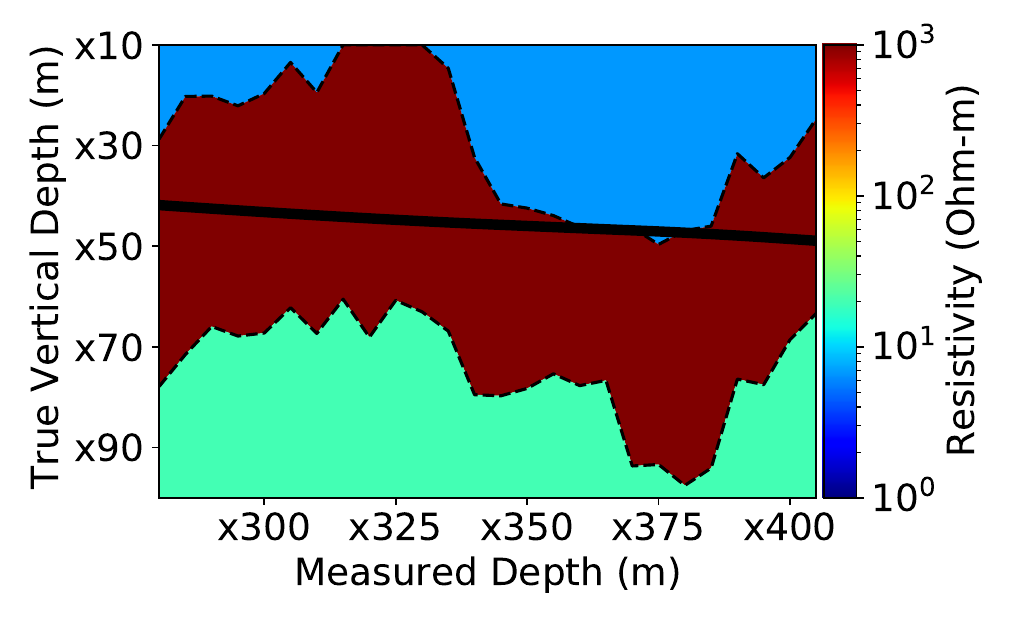}	
	\end{subfigure}
 \hspace{0.45in}
	\begin{subfigure}[normal]{0.4\textwidth}
	\includegraphics[scale=0.5]{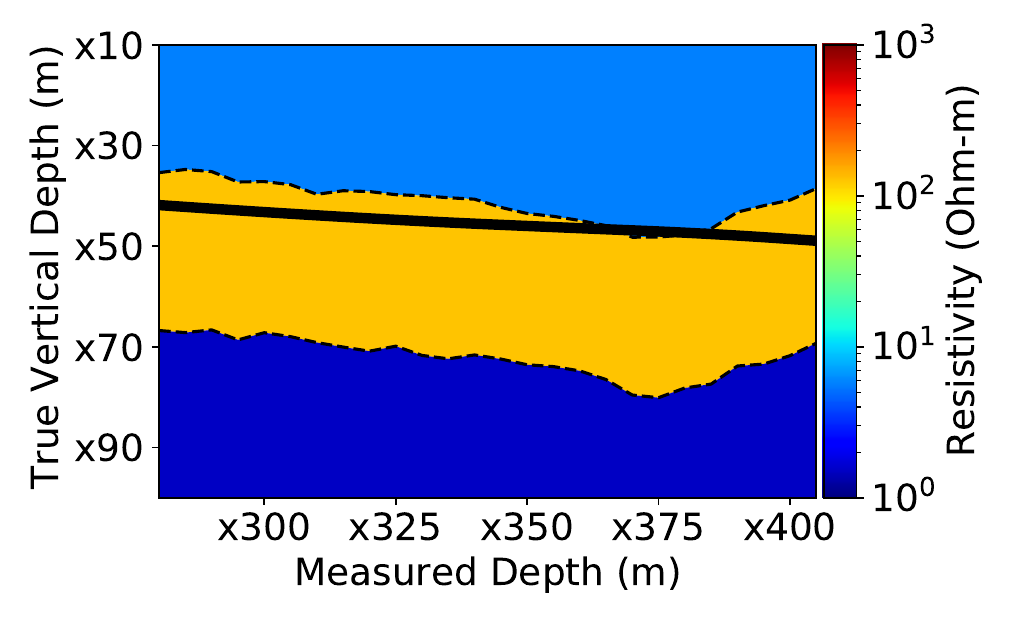}	
	\end{subfigure}

 \hspace{0.4in}
		 ESMDA (neglecting model-error)
 \hspace{0.7in}
 		 FlexIES (accounting for model-error)
 \end{center}

\caption{Estimation results of the geo-layers profile including the logs with large model errors.}
\label{est_4a}
\end{figure}

Assimilating such data into a classical IES results erroneous updates, see Figure \ref{est_4a} left (only P50 shown). 
Without a prior analysis of the output data and manual filtering, as done in Case 1, assimilating such measurements can result in biased updates and sub-optimal decisions.

\begin{figure}
\begin{center}

 \hspace{-0.5in}
    \begin{subfigure}[normal]{0.4\textwidth}
	\includegraphics[scale=0.5]{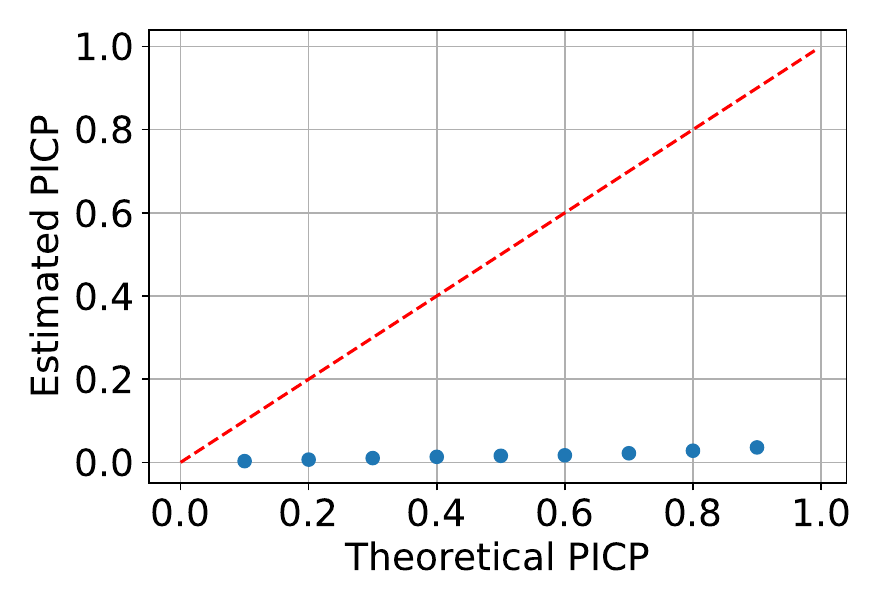}	
	\end{subfigure}
 \hspace{0.5in}
	\begin{subfigure}[normal]{0.4\textwidth}
	\includegraphics[scale=0.5]{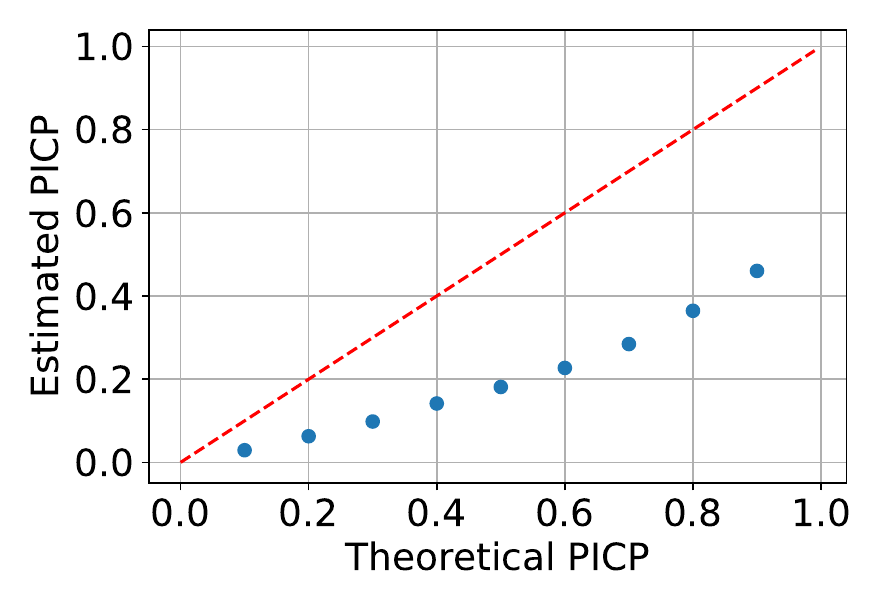}	
	\end{subfigure}

 \hspace{0.4in}
		 ESMDA (neglecting model-error)
 \hspace{0.7in}
 		 FlexIES (accounting for model-error)		 
 		 
\end{center} 
 
\caption{PICP after assimilating measurements with large model error.}
\label{picp_3}
\end{figure}

\begin{figure}
\begin{center}

 \hspace{-0.5in}
    \begin{subfigure}[normal]{0.4\textwidth}
	\includegraphics[scale=0.5]{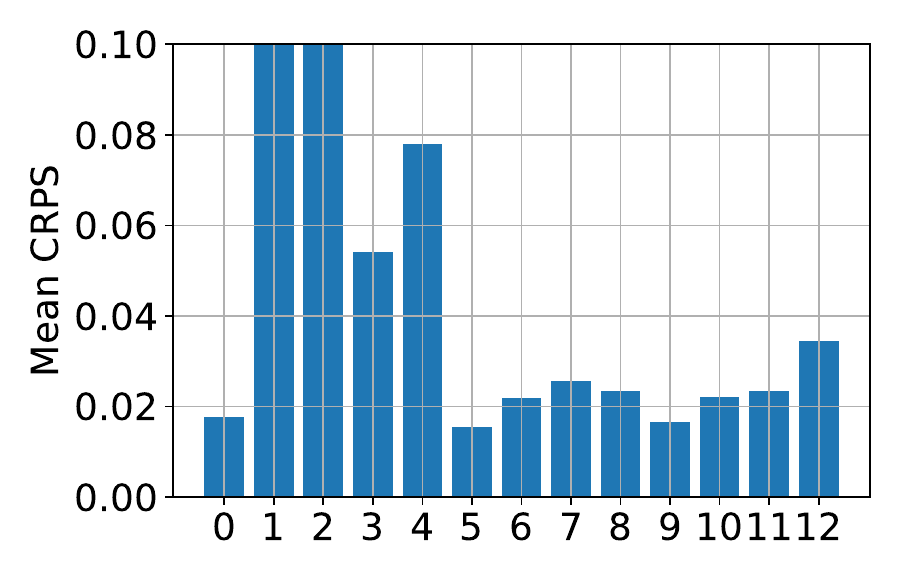}	
	\end{subfigure}
 \hspace{0.5in}
	\begin{subfigure}[normal]{0.4\textwidth}
	\includegraphics[scale=0.5]{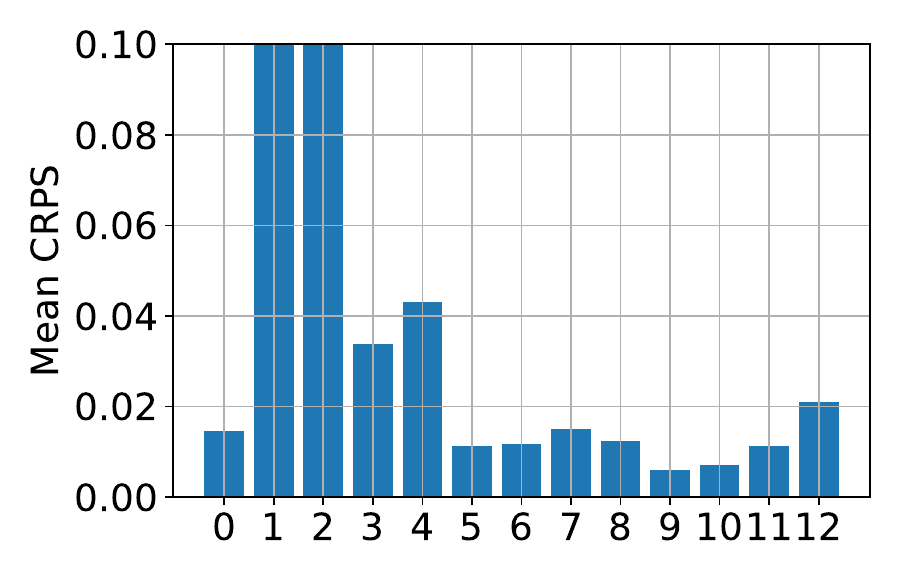}	
	\end{subfigure}

 \hspace{0.4in}
		 ESMDA (neglecting model-error)
 \hspace{0.7in}
 		 FlexIES (accounting for model-error)		 
 		 
\end{center} 
 
\caption{Mean CRPS after assimilating measurements with large model error. X-axis shows the log numbers.}
\label{crps_3}
\end{figure}

At the same time, by automatic detection of model errors the FlexIES algorithm manages to ignore the inconsistencies of these shallow logs and provides a stable solution similar to the case 1, where the shallow logs were manually removed, see Figure \ref{est_4a} right. The PICP metrics after inverting all outputs data show relatively poor performance as compared to the previous cases as shown in Figure \ref{picp_3}. 
Even though the PICP values when using all the output data becomes low using FlexIES as compared to previous cases, it is still usable, unlike the traditional IES method.
The effect of the model error results in the high values of the mean CRPS metric of all logs as shown in Figure \ref{crps_3}. However, FlexIES lowers the mean CRPS values thus improves the results of all logs while accounting for model error except logs 1 and 2, see Figure \ref{crps_3} (right panel).

\section{Conclusions}   
\label{sec:conclusions}

In this paper, we demonstrated a new workflow for real-time estimation of uncertain geo-layer profiles for field operations using a fast and approximate DNN model of the EDAR measurements. 
The workflow consists of offline (preparation) and online phases.
In the offline phase, one constructs a relevant geological prior and trains the DNN.

The online phase uses a trained DNN as a forward model in the ensemble-based FlexIES data assimilation loop, reducing uncertainties while accounting for model errors. 

We demonstrated the performance of this workflow on a section of a historical operation from the Goliat field. We showed that the Bayesian interpretation could be applied in seconds to estimate the geo-layer profiles and layer resistivities in an anisotropic environment. 
Thus, the DNN model enables costly ensemble-based IES workflows in real time.
However, we observed that classical IES, which neglects model errors during the interpretation, often provides erroneous estimates of the subsurface, which in practice will yield suboptimal decisions.
In these cases, FlexIES provides reliable estimates of the geo-layer profiles and uncertainty in complex, realistic conditions, even with an approximate model. 
The median of our probabilistic interpretation is on-par with the proprietary deterministic inversion. 
Moreover, the method estimates the uncertainties in the geo-model parameters, which can be very useful for real-time decision support.

We described the minor modifications required for the combined framework of data-driven deep learning models and FlexIES implementation in field operations.
The main benefit will come from the integration with decision support, e.g. using dynamic programming \cite{alyaev2019decision}.
In the future, this work can be extended and applicable to account for model errors in more complex geo-models by, e.g. using generative adversarial networks \cite{alyaev2021probabilistic,fossum2022verification}.
Furthermore, the proposed workflow gives a complete framework to assimilate different types of measurements with possibly approximate, less accurate, and low-fidelity surrogate models.

\section{Acknowledgments}
We thank the Goliat license team from Vår Energi for sharing the data.

The work was started as part of  the research project ’Geosteering for
IOR’ (NFR-Petromaks2 project no. 268122) funded by the Research
Council of Norway, Aker BP, Equinor, V{\aa}r Energi and Baker Hughes Norway.

Muzammil Hussain Rammay was supported by the Equinor's Academia fund and the research project 'Geosteering for IOR'.

Sergey Alyaev was supported by the Center for Research-based Innovation DigiWells: Digital Well Center for Value Creation, Competitiveness and Minimum Environmental Footprint (NFR SFI project no. 309589, https://DigiWells.no). The center is a cooperation of NORCE Norwegian Research Centre, the University of Stavanger, the Norwegian University of Science and Technology (NTNU), and the University of Bergen. It is funded by Aker BP, ConocoPhillips, Equinor, Lundin Energy, TotalEnergies, Vår Energi, Wintershall Dea, and the Research Council of Norway.

\begin{appendices}

\section{Evaluation of the trained DNN}
\label{sec:trainingResults}

\begin{figure}
    \centering
    \includegraphics[width=0.45\textwidth]{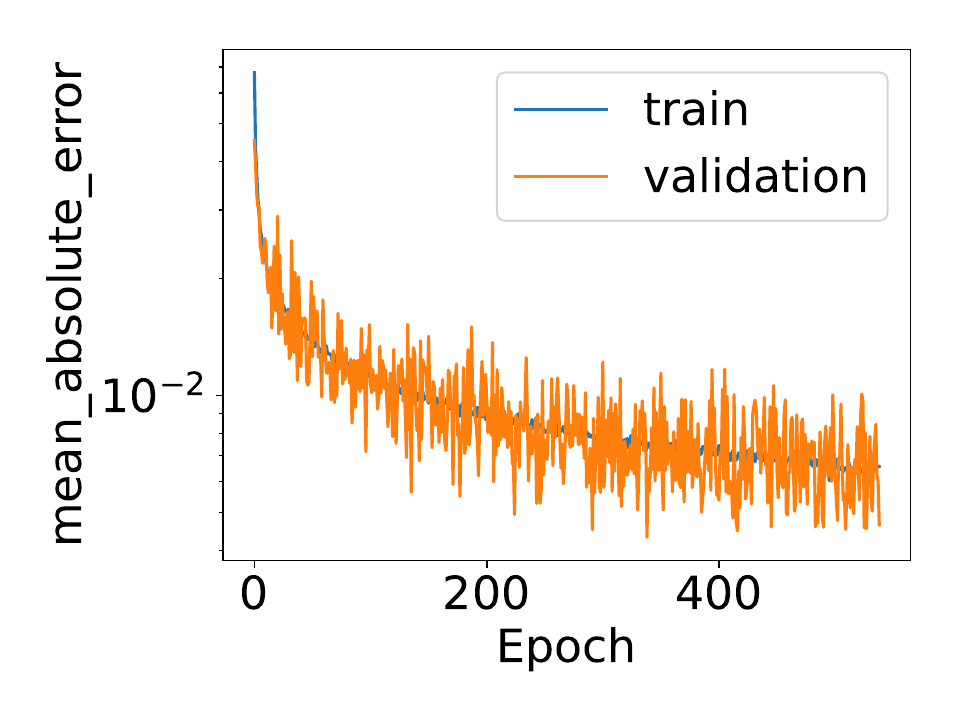}
    \caption{Evolution of the mean-absolute-error loss training and validation datasets during the training of the DNN.}
    \label{fig:dnn-taraining}
\end{figure}

Due to a larger and more expressive dataset compared to \cite{alyaev2021modeling} we used smaller value of 200 epochs for early stopping.
During our training it was triggered after about 538 epochs. 
The evolution of training and validation losses are shown in Figure \ref{fig:dnn-taraining}. The error decreases steadily for both training and validation, and the final loss magnitudes are comparable with results for a simpler dataset described in \cite{alyaev2021modeling}. 

\begin{table}[]
    \centering
    \includegraphics[width=0.7\textwidth]{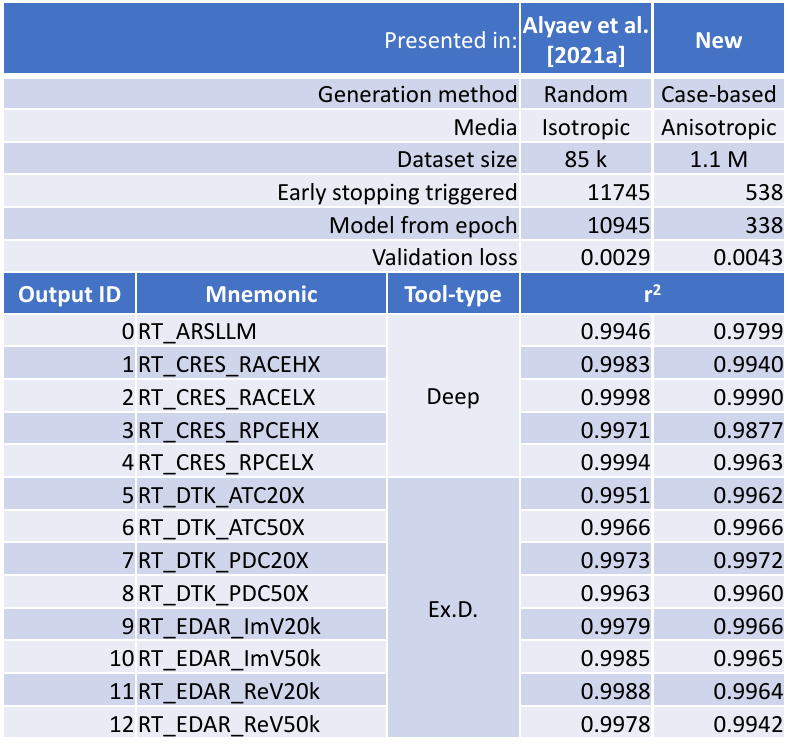}
    \caption{Comparison of the training setups and the testing results between \cite{alyaev2021modeling} and the current paper. $r^2$ is the coefficient of determination between the reference values from the test dataset unseen by the DNN and the DNN predictions; 1.0 is the perfect fit.}
    \label{tab:training}
\end{table}

Table \ref{tab:training} gives a detailed overview of the training setup and results and compares them to the paper where the DNN architecture was proposed \citep{alyaev2021modeling}. While the validation loss is higher for the new model, the coefficients of determination for the test data, never seen by the model, are comparable between the old randomly generated and the new field-case-based test cases. We observe fractionally worse qualities for outputs number 0 and 3.

\begin{figure}
    \centering
    \includegraphics[width=0.45\textwidth]{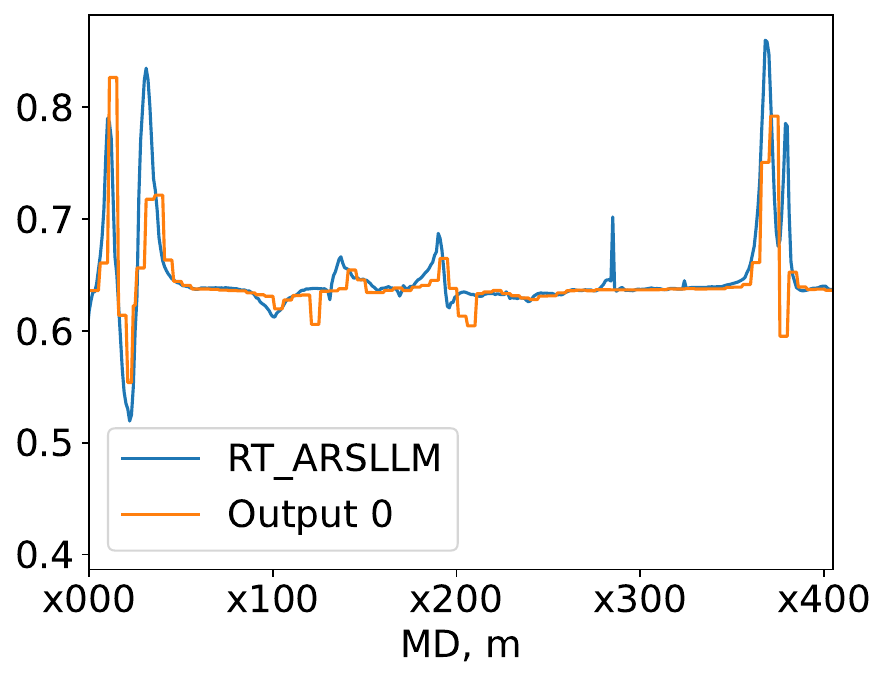}\\
    \includegraphics[width=0.45\textwidth]{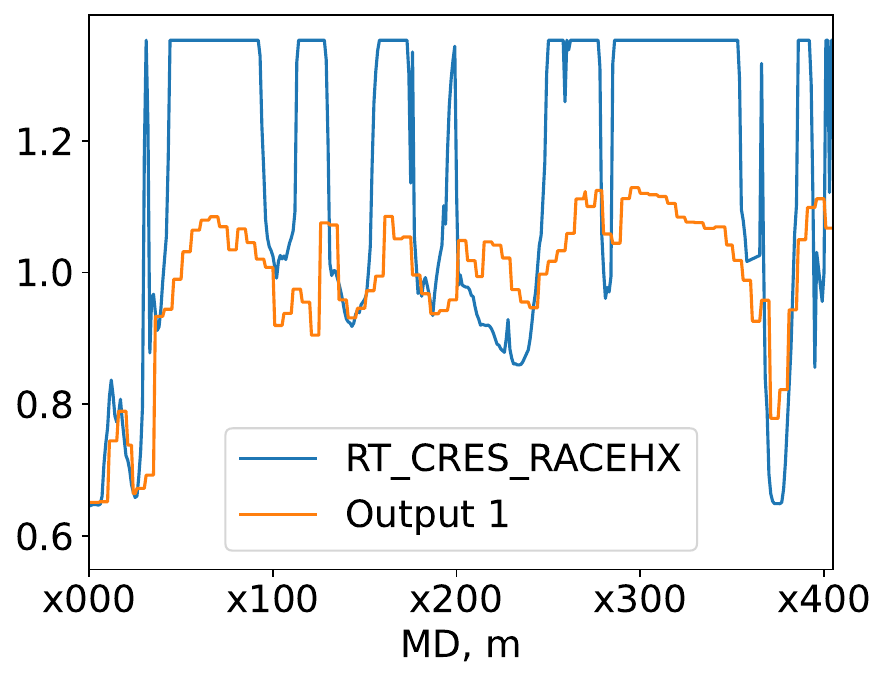}
    \includegraphics[width=0.45\textwidth]{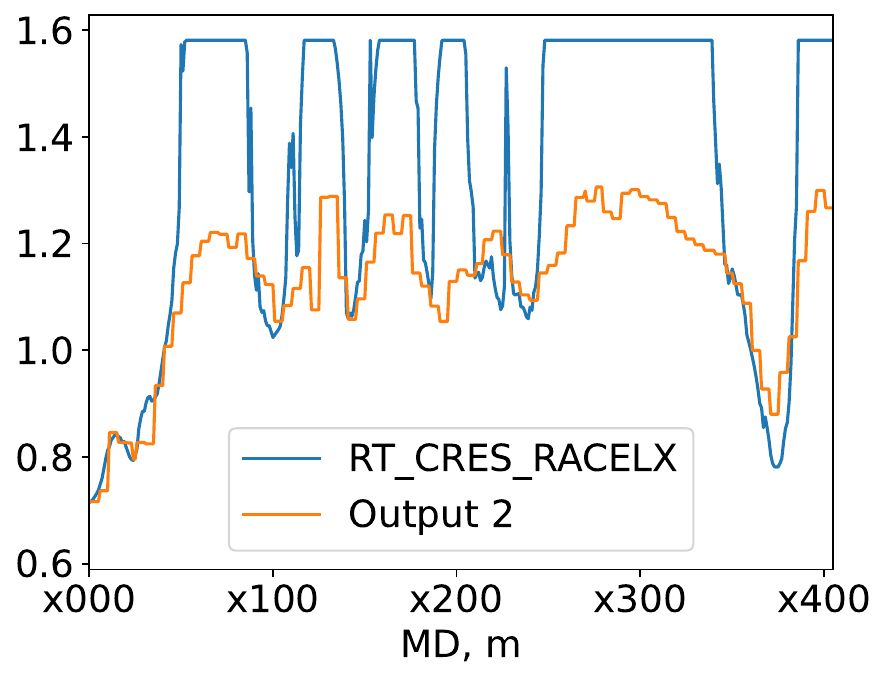}\\
    \includegraphics[width=0.45\textwidth]{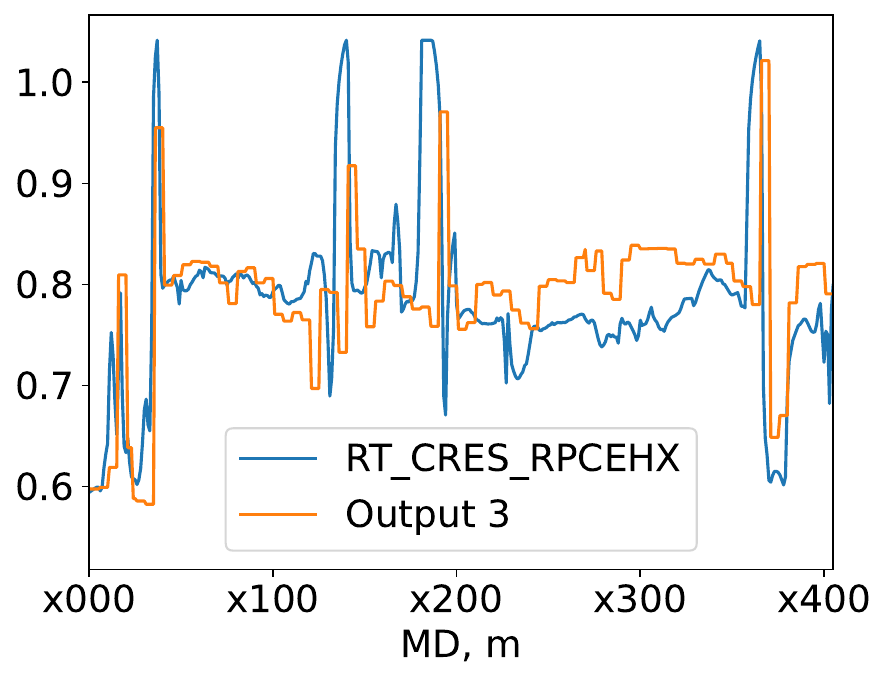}
    \includegraphics[width=0.45\textwidth]{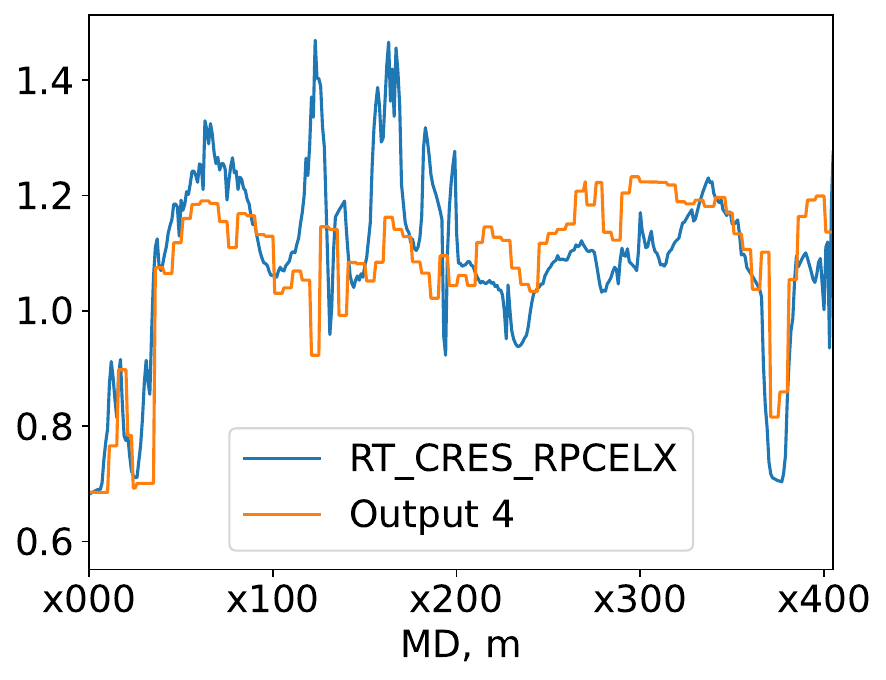}
    \caption{The deep EM real-time data from the Goliat field compared to the
    logs modeled by DNN from a reference inversion model presented in \cite{larsen2015extra}. All curves are re-scaled to 0.5..1.5 using the scaling introduced in Section~\ref{sec:logTypess}.}
    \label{fig:realLogsDeep}
\end{figure}

\begin{figure}
    \centering
    \includegraphics[width=0.45\textwidth]{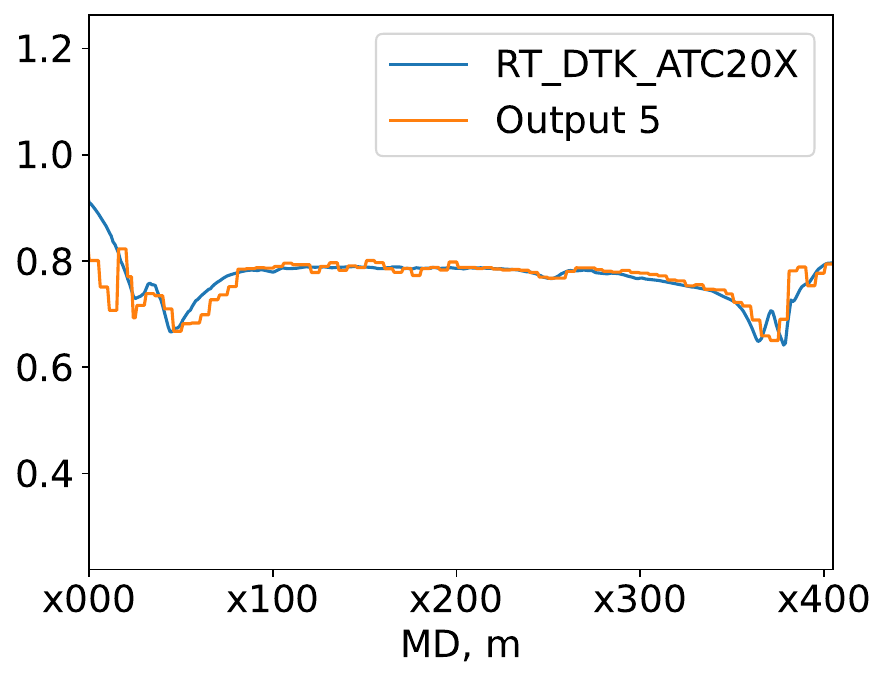}
    \includegraphics[width=0.45\textwidth]{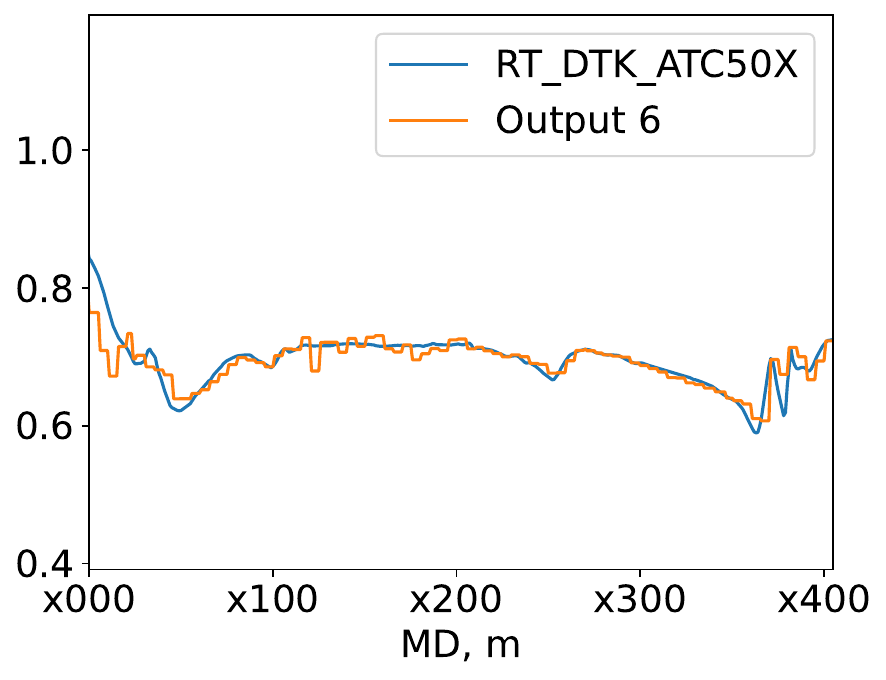}\\
    \includegraphics[width=0.45\textwidth]{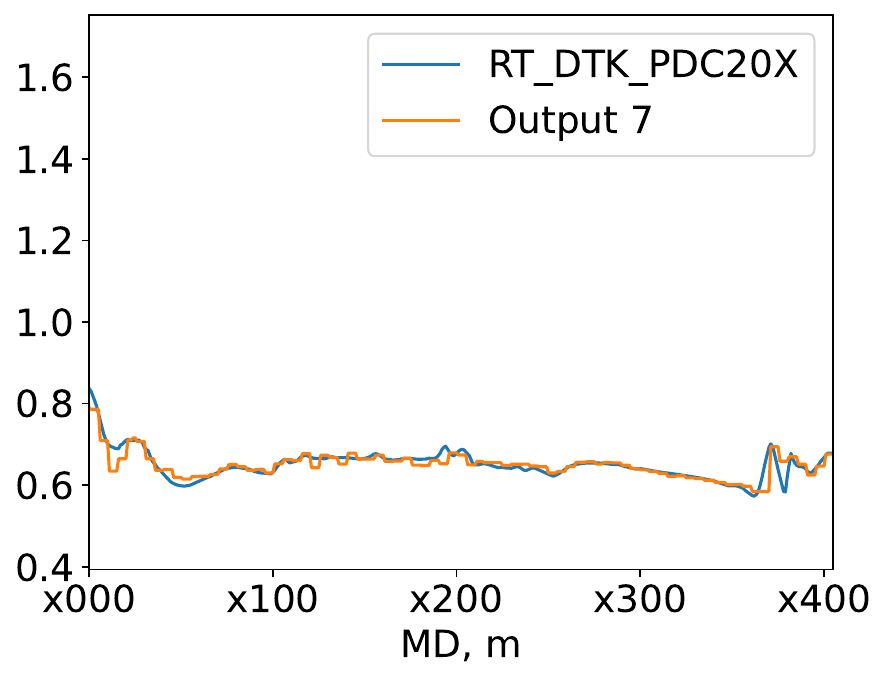}
    \includegraphics[width=0.45\textwidth]{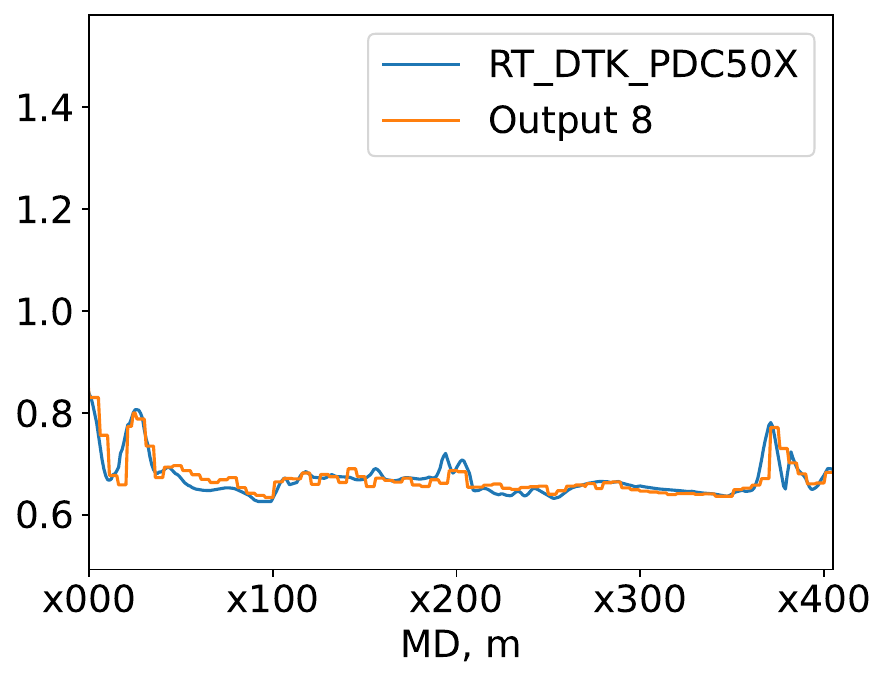}\\
    \includegraphics[width=0.45\textwidth]{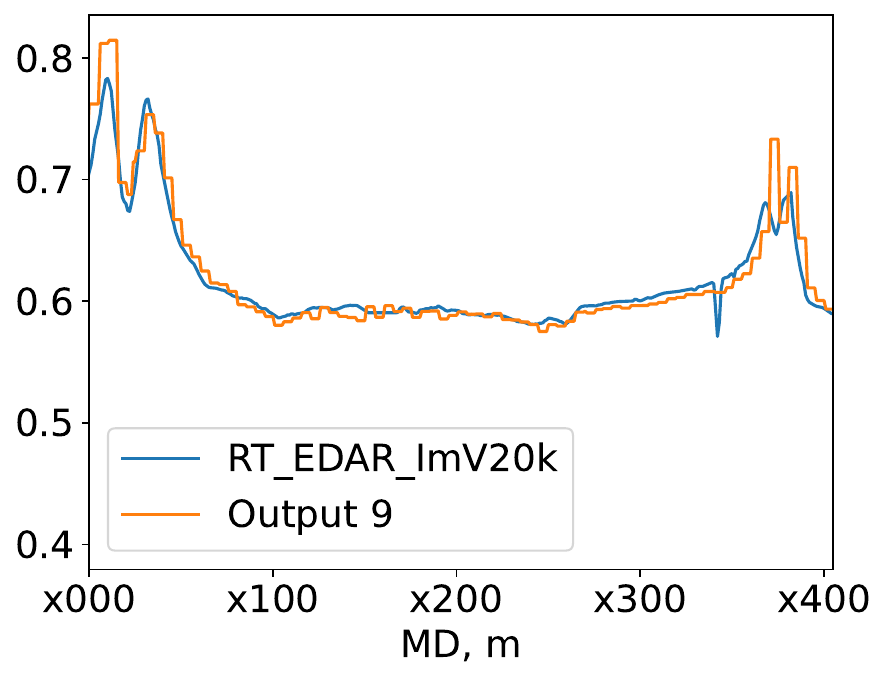}
    \includegraphics[width=0.45\textwidth]{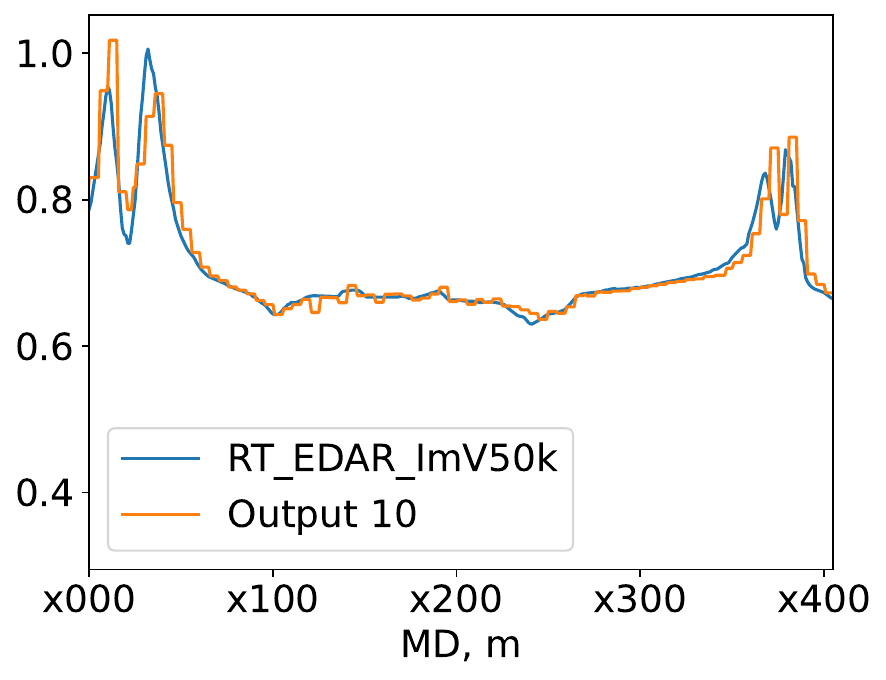}\\
    \includegraphics[width=0.45\textwidth]{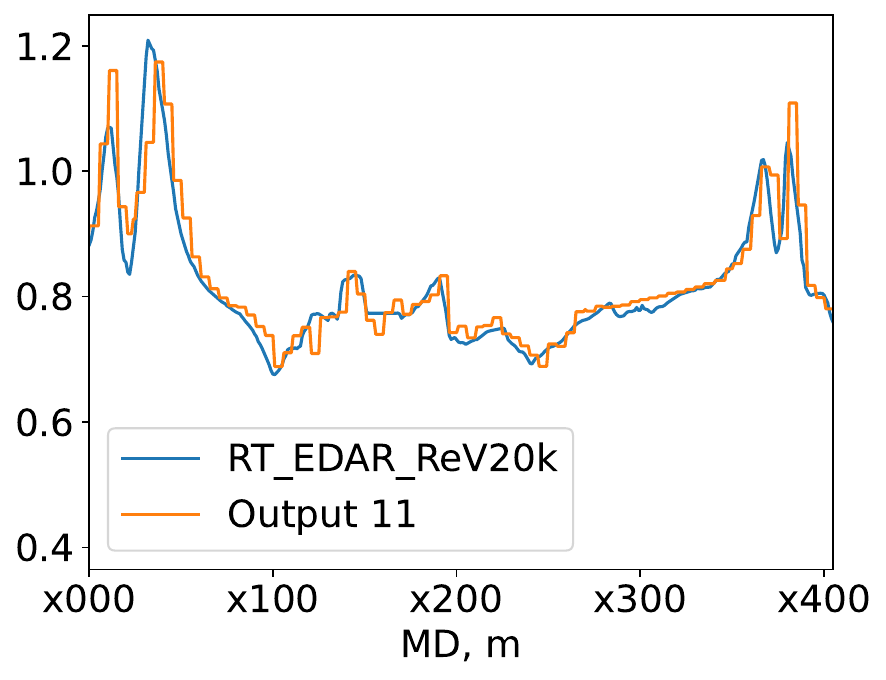}
    \includegraphics[width=0.45\textwidth]{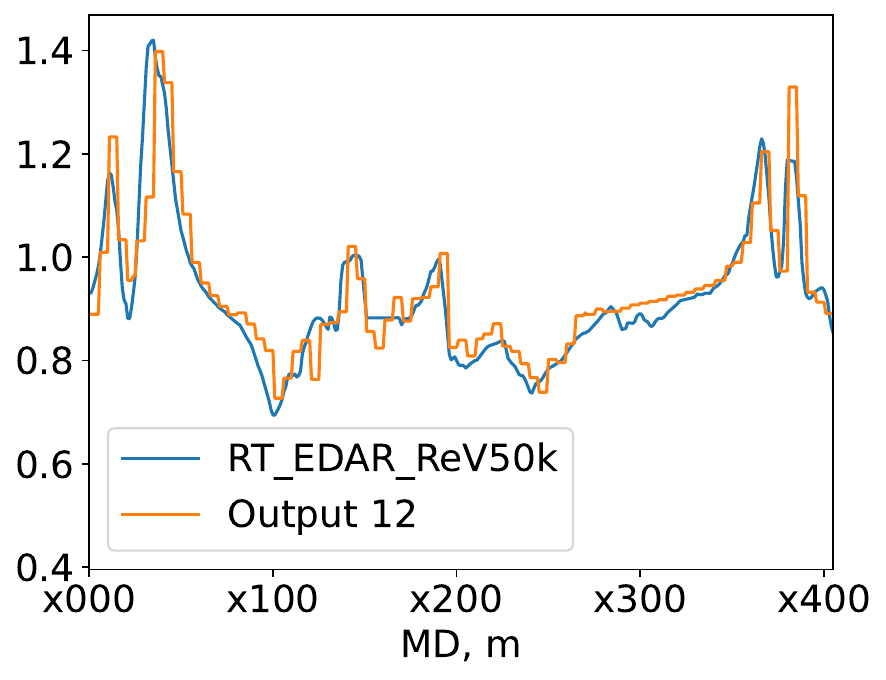}
    \caption{The extra-deep EM (EDAR) real-time data from the Goliat field compared to the
    logs modeled by DNN from a reference inversion model presented in \cite{larsen2015extra}. All curves are re-scaled to 0.5..1.5 using the scaling introduced in Section~\ref{sec:logTypess}}
    \label{fig:realLogsExtraDeep}
\end{figure}

Equipped with the trained DNN it is of interest to compare its predictions to actual logs from the studied field.
We apply the trained DNN model to the vendor-provided inversion of the field data from \cite{larsen2015extra}, shown in Figure \ref{prop_inversion}.
Figures \ref{fig:realLogsDeep} and \ref{fig:realLogsExtraDeep} show the differences between the real data and the data modelled by the DNN (all scaled).
While the DNN model matches the synthetic data from the test dataset really well, there are pronounced misfits when comparing to the field data.
This can be caused by model errors unaccounted in the physics-based model, such as thin beddings, hetrogeneities etc. The shallow-sensing outputs 1 and 2 (Figure \ref{fig:realLogsDeep}) are specifically susceptible to near-well variations and show big differences to the model. 
Therefore they are left out from most of our inversion experiments.

\section{Pseudocode of the FlexIES} \label{sec:AppendixA}
\setcounter{equation}{0}
\renewcommand{\theequation}{\thesection.\arabic{equation}}
\renewcommand{\thetable}{\thesection.\arabic{table}}
\def\NoNumber#1{{\def\alglinenumber##1{}\State #1}\addtocounter{ALG@line}{-1}}

\begin{algorithm}
\renewcommand{\thealgorithm}{}
\caption{FlexIES function for EM inversion using DNN model $\rhd$ \citep{rammay2020flexible}
}
\begin{algorithmic}

\State

\Function {FlexIES}{$\mathbf{M}$, $\mathbf{EM}_{obs}$, $\mathbf{C}_D$, $DNN$, $\mathbf{t}$} 
\State \textbf{Inputs:} $\mathbf{M} \in \mathbb{R}^{N_m \times N_e}$ is the ensemble of input parameters, $\mathbf{C}_D$ is the covariance matrix of the measurement error, $\mathbf{t}$ is the given well trajectory.
\State Choose $N_a$ \quad $\rhd$ Number of data assimilations/iterations
\State $ i \gets 1$
\State $\alpha = N_a$

\While {$ i <= N_a$ }
\State $\mathbf{D}_{uc} = \mathbf{EM}_{obs}\mathbf{\overrightarrow{1_{N_e}}} + \sqrt{\alpha}\mathbf{C}_D^{1/2}\mathbf{Z}_d$ $\ \ \rhd$ Observation perturbations, $\mathbf{Z}_{d} = [\mathbf{z}_{d1}\; \mathbf{z}_{d2} \; \mathbf{z}_{d3} \; ...... \; \mathbf{z}_{dN_e}]$ $\in\mathbb{R}^{N_d \times N_e}$, $\mathbf{z}_d \sim \mathcal{N}(0, \mathbf{I}_{N_d}) \in \mathbb{R}^{N_d \times 1}$, ${\mathbf{\overrightarrow{1_{N_e}}} \in \mathbb{R}^{1 \times N_e}}$ is a row vector of ones.

\State $ \mathbf{D} = DNN(\mathbf{M,t}) $  \quad  $\rhd$ Generate ensemble of model outputs $\mathbf{D} \in \mathbb{R}^{N_d \times N_e}$ from $\mathbf{M}$

\State $ \mathbf{R}  = \mathbf{EM}_{obs} \mathbf{\overrightarrow{1_{N_e}}} - \mathbf{D}$

\If {$i = 1$}
\State ${s_p}^{(i)} = \frac{\|{\boldsymbol{\sigma}_{m}}^{(i)}\|}{\|\boldsymbol{\sigma}_{max}\|}$   \quad $\rhd$ Compute split parameter, $\boldsymbol{\sigma}_{max} = \operatorname{max}(\operatorname{abs}(\mathbf{R})) \in \mathbb{R}^{N_d \times 1}$
\Else
\State ${s_p}^{(i)} = \frac{\|{\boldsymbol{\sigma}_{m}}^{(i)}\|}{\|{\boldsymbol{\sigma}_{m}}^{(i-1)}\|}$  \quad $\rhd$ $||.||$ is the Euclidean (L2) norm , $\boldsymbol{\sigma}_{m} = \operatorname{mean}(\mathbf{R}) \in \mathbb{R}^{N_d \times 1}$ 
\EndIf

\State $\mathbf{E} =  {s_p}^{(i)} \mathbf{R}$ \quad  $\rhd$ Compute ensemble of approximate model-error, $\mathbf{E}  \in \mathbb{R}^{N_d \times N_e}$

\State	$\mathbf{C}_{EE} = \frac{1}{N_e - 1} (\mathbf{E}-\mathbf{\bar{E}} \mathbf{\overrightarrow{1_{N_e}}})(\mathbf{E}-\mathbf{\bar{E}} \mathbf{\overrightarrow{1_{N_e}}})^{^\intercal}$ \quad $\rhd$ $\mathbf{\bar{E}} \in \mathbb{R}^{N_d \times 1}$ is ensemble mean of $\mathbf{E}$

\State $\mathbf{M}  \gets \mathbf{M} + \mathbf{C}_{MD} ~(\mathbf{C}_{DD} + \mathbf{C}_{EE} + \alpha~\mathbf{C}_D)^{-1}(\mathbf{D}_{uc}-\mathbf{D} - \mathbf{E})$  \quad $\rhd$ Update ensemble $\mathbf{M}$

\State $ i \gets i + 1 $
\EndWhile

\State $\mathbf{M}_{post} = \mathbf{M}$ \quad $\rhd$ $\mathbf{M}_{post}  \in \mathbb{R}^{N_m \times N_e}$ is the posterior ensemble of input parameters

\State $\mathbf{D}_{post} = DNN(\mathbf{M}_{post},\mathbf{t}) $ \quad $\rhd$ Generate posterior ensemble of model outputs, $\mathbf{D}_{post}$
\State \textbf{return} $\mathbf{M}_{post}$, $\mathbf{D}_{post}$
\EndFunction

\end{algorithmic}
\end{algorithm}

\section{Inversion assessment metrics}	\label{sec:AppendixB}
\setcounter{equation}{0}
\renewcommand{\theequation}{\thesection.\arabic{equation}}
\renewcommand{\thetable}{\thesection.\arabic{table}}

\subsection{Prediction interval coverage probability (PICP)}

The mathematical description of Coverage Probability (CP) is shown below,
\begin{equation}
CP = \frac{N_{CI}}{N_t}.
\end{equation}

$N_{CI}$ = Number of observations or parameters in Confidence Interval

$N_t$ = Total number of observations or parameters

Prediction interval coverage probability is obtained by computing coverage probability for confidence interval $10\%, 20\%, ..., 80\%, 90\%$.

\subsection{Continuous Ranked Probability Score (CRPS)}

Mathematically, CRPS can be described as follows, \citep{hersbach2000decomposition}

\begin{equation}
CRPS = \int_{-\infty}^{\infty}[p(x) - H(x-x_{obs})]^2 dx,
\end{equation}
where $p(x) = \int_{-\infty}^{x} \rho(y) dy$  Cumulative distribution of quantity of interest, $H(x-x_{obs})$ = Heaviside function (Step function) i.e.
\[
  H(x) =
  \begin{cases}
   0            & \text{if $x<0$} \\
   1            & \text{if $x\geq0$} \\
  \end{cases}
\]

For an ensemble system with $N_e$ realizations, the CRPS can be written as follows,
\begin{equation}
CRPS = \sum_{r=0}^{N_e}{c_r}.
\end{equation}

\begin{equation}
c_r = \alpha_r p_r^2 + \beta_r (1-p_r)^2.
\end{equation}
where $p_r = P(x) = r/N_e$,   for $x_r < x < x_{r+1}$  (Cumulative distribution is a piecewise constant function).
\[
  \alpha_r =
  \begin{cases}
  0                                  & \text{if $x_{obs}<x_r$} \\
  x_{obs} - x_{r}                    & \text{if $x_r<x_{obs}<x_{r+1}$} \\
  x_{r+1} - x_{r}                    & \text{if $x_{obs}>x_{r+1}$}   \\
  x_{obs} - x_{N_e}                  & \text{if $x_{obs}>x_{N_e}$}   \\
  0                                  & \text{if $x_{obs}<x_1$}   \\
  \end{cases}
\]

\[
  \beta_r =
  \begin{cases}
  x_{r+1} - x_{r}                       & \text{if $x_{obs}<x_r$} \\
  x_{r+1} - x_{obs}                     & \text{if $x_r<x_{obs}<x_{r+1}$} \\
  0                                     & \text{if $x_{obs}>x_{r+1}$}   \\
  0                                     & \text{if $x_{obs}>x_{N_e}$}   \\
  x_1 - x_{obs}                         & \text{if $x_{obs}<x_1$}   \\
  \end{cases}
\]

\end{appendices}

\bibliographystyle{apalike}
\bibliography{bibfile2}

\end{document}